\documentclass{article}       
\usepackage{graphicx}
\usepackage{color,verbatim}
\usepackage{lscape}
\usepackage{rotating}
\usepackage{wrapfig}
\usepackage{multirow}
\usepackage{placeins}
\usepackage{amsmath,bm,natbib}
\usepackage{amsfonts}
\usepackage{amssymb}
\usepackage{graphicx,psfrag,epsf,caption,subcaption,float}
\usepackage{enumerate, enumitem}
\usepackage{url} 
\usepackage{statex2}
\usepackage[font=small,labelfont=bf]{caption}
\usepackage{adjustbox}
\usepackage{multirow}

\newcommand{\BART}{\mathrm{BART}}
\newcommand{\bx}{\bm{x}}

\newcommand{\cn}{\mathrm{N}}
\newcommand{\ct}{\mathcal{T}}
\newcommand{\cm}{\mathcal{M}}
\newcommand{\expit}{\mathrm{expit}}

\begin{document}

\title{Dynamic Treatment Regimes using Bayesian Additive Regression Trees for Censored Outcomes}

\author{Xiao Li, Brent R Logan, S M Ferdous Hossain, and Erica E M Moodie}

\maketitle




\begin{abstract}
To achieve the goal of providing the best possible care to each patient, physicians need to customize treatments for patients with the same diagnosis, especially when treating diseases that can progress further and require additional treatments, such as cancer. Making decisions at multiple stages as a disease progresses can be formalized as a dynamic treatment regime (DTR). Most of the existing optimization approaches for estimating dynamic treatment regimes including the popular method of Q-learning were developed in a frequentist context. Recently, a general Bayesian machine learning framework that facilitates using Bayesian regression modeling to optimize DTRs has been proposed. In this article, we adapt this approach to censored outcomes using Bayesian additive regression trees (BART) for each stage under the accelerated failure time modeling framework, along with simulation studies and a real data example that compare the proposed approach with Q-learning. We also develop an R wrapper function that utilizes a standard BART survival model 
to optimize DTRs for censored outcomes. The wrapper function can easily be extended to accommodate any type of Bayesian machine learning model.  \\ \ \\

\end{abstract}

\section{Introduction}
\label{sec:intro}
Optimizing medical therapy often requires that the treatment be individually tailored to the patient initially, and that the treatment be adaptive to patient's changing characteristics over time. Since patient responses can often be heterogeneous, it is challenging for physicians to customize treatments for patients based on traditional clinical trial results, which lack the ability to identify subgroups that have different treatment effects and rarely consider successions of treatments. For chronic diseases that can evolve, it is even more important and difficult to choose the best therapy in sequence. To give a simple example, oncologists typically choose an initial immunosuppressant regime for patients with acute myeloid leukemia (AML) who are undergoing allogeneic hematopoietic cell transplantation (AHCT), to prevent a serious potential complication called graft-versus-host disease (GVHD). At the time that such an initial regime fails, a salvage treatment is chosen based on the patient’s prior treatments and responses. Such a multi-stage treatment decision has been summarized as a dynamic treatment regime (DTR) by Murphy \cite{murphy2003optimal}. Each decision rule in DTR takes a patient's individual characteristics, treatment history and possible intermediate outcomes observed up to a certain stage as inputs, and outputs a recommended treatment for that stage.

A number of approaches have been proposed for estimating and optimizing DTRs, including those by Robins  \cite{robins2004optimal}, Moodie et al.\  \cite{moodie2007demystifying}, Qian and Murphy  \cite{qian2011performance}, Zhao et al.\  \cite{zhao2015new}, Krakow et al.\  \cite{krakow2017tools}, Murray et al.\  \cite{murray2018bayesian}, and Simoneau et al.  \cite{simoneau2020estimating}. Among the previous literature, the Bayesian machine learning (BML) method developed by Murray et al.\ \cite{murray2018bayesian} innovatively bridges the gap between Bayesian inferences and dynamic programming methods from machine learning. A key advantage to a Bayesian approach to estimation is the quantification of uncertainty in decision making through the resulting posterior distribution. A second benefit that arises specifically in the BML approach is the highly flexible estimation that is employed, which minimizes the risk of estimation errors due to model mis-specification. 

However, the BML method has not yet been adapted to censored outcomes, which is one of the more common types of outcomes in controlling chronic diseases. Motivated by the study of optimal therapeutic choices to prevent and treat GVHD, in this paper, we extend this approach to censored outcomes under the accelerated failure time (AFT) model framework. By modifying the data augmentation step in the BML method, the censored observations can be imputed in an informative way so that the observed censoring time is well utilized. This extension is illustrated using Bayesian additive regression trees (BART). We also implemented the proposed AFT-BML approach by developing an R function that utilizes standard BART survival software directly without needing to modify existing (complex) BART software directly.  Parallel computing was used to speed up the computational calculations. This R wrapper function can be easily adjusted to accommodate other types of Bayesian machine learning methods.

This paper is organized as follows. In Section \ref{sec:meth}, we briefly review related methods, algorithms, and describe the extended AFT-BML approach for optimizing DTRs. Section \ref{sec:simu} presents simulation studies to demonstrate our model performance by comparing to it to estimation using Q-learning. An analysis of our motivating dataset  of patients diagnosed with AML is given in Section \ref{sec:ex}. Finally, in Section \ref{sec:conc} we discuss the advantages and disadvantages of our approach and provide some suggestions for future work.

\section{Methods}
\label{sec:meth}
\subsection{Dynamic Treatment Regimes}
A dynamic treatment regime (DTR) is a series of decision rules that assign treatment based on the patient's characteristics and history at each stage. Without loss of generality, we focus on a two-stage intervention problem. Furthermore, we start by describing DTRs in the non-survival setting, before proceeding to the censored survival setting later.  Following Murray's notation  \cite{murray2018bayesian}, let $o_1\in\mathcal{O}_1$ be the covariates observed before Stage 1, and $a_1\in\mathcal{A}_1$ be the action taken at Stage 1. Denote $y_1$ as the payoff observed after Stage 1 and before Stage 2. $\{o_2,a_2,y_2\}$ are defined similarly for Stage 2. The total payoff is assumed to be $y=y_1+\eta y_2$, where $\eta$ is an indicator that the patient entered Stage 2. A general diagram to present the two-stage decision making problem is
\[o_1 \longrightarrow a_1 \longrightarrow y_1 \overset{\mbox{ if }\eta=1}{\longrightarrow} o_2 \longrightarrow a_2 \longrightarrow y_2.\]
Denote the accumulated history before Stage 2 treatment as $\bar{o}_2=(o_1,a_1,y_1,o_2)\in\mathcal{\bar{O}}_2$. In this setting, a DTR consists of two decision rules, one for each stage,
\[d_1:\mathcal{O}_1\rightarrow \mathcal{A}_1\quad \mbox{ and }\quad d_2:\mathcal{\bar{O}}_2\rightarrow \mathcal{A}_2.\]
Optimizing the two-stage DTR $(d_1,d_2)$ is equivalent to finding the decision rules $d_1$ and $d_2$ that maximize the expected total payoff $E(y)$. That is
\begin{align*}
    d_2^{opt}(\bar{o}_2) &= \arg \sup_{a_2\in\mathcal{A}_2} E(y_2|\bar{o}_2, a_2) & \forall \bar{o}_2\in \mathcal{\bar{O}}_2, \\
    d_1^{opt}(o_1) &= \arg \sup_{a_1\in\mathcal{A}_1} E(y|o_1,a_1, d_2^{opt}) & \forall o_1\in\mathcal{O}_1.
\end{align*}

\subsection{Bayesian Machine Learning for DTRs}
Murray et al.\   \cite{murray2018bayesian} described a new approach called Bayesian Machine Learning (BML) to optimize DTRs; the method requires fitting a series of Bayesian regression models in reverse sequential order under the approximate dynamic programming framework. The authors use the potential outcomes notation to describe their approach, where $y(a_1,a_2)$ denotes the payoff observed when action $a_1$ is taken at Stage 1 and action $a_2$ is taken at Stage 2, and other potential outcomes ($y_2(a_1,a_2)$, $y_1(a_1)$, and $o_2(a_1)$) are similarly defined.  Assuming the potential outcomes are consistent, then the observed outcome corresponds to the potential outcome for the action actually followed, e.g.  $y_1(a_1)=y_1$, $o_2(a_1)=o_2$,  $y_2(a_1,a_2)=y_2$, and $y(a_1,a_2)=y$.  The approach can be summarized as follows. The Stage 2 regression model for $y_2(a_1,a_2)$ is estimated first, using the observed covariates $(\bar{o}_2, a_2)$ and the observed response variable $y_2$. Based on the estimated Stage 2 model, the estimated optimal mapping from $\mathcal{\bar{O}}_2$ to $\mathcal{A}_2$, simply denoted as $d_2^{opt}$, can be identified, as well as the relevant potential payoff at Stage 2, denoted as $y_2(a_1,d_2^{opt})$. With $d_2^{opt}$ and potential payoff $y_2(a_1,d_2^{opt})$, the response variable for Stage 1 can be constructed as $y(a_1,d_2^{opt})$; this (potential) outcome is composed of the observed Stage 1 payoff $y_1$ and the potential Stage 2 payoff $y_2(a_1,d_2^{opt})$. Note that if the observed outcome $a_2$ matches the optimal outcome according to $d_2^{opt}$, then the potential payoff is simply the observed payoff $y=y_1+\eta y_2$.  Otherwise, the potential outcome is unobserved and must be imputed (in this BML method, actually sampled from the posterior predictive distribution as described further below).  Given imputed values, the Stage 1 regression model for $y(a_1,d_2^{opt})$ then can be estimated with observed covariates $(o_1,a_1)$ to identify $d_1^{opt}$. This type of backward induction strategy is used in several DTR estimation methods, including g-estimation, Q-learning, dynamic weighted ordinary least squares \cite{robins2004optimal,moodie2007demystifying,NahumShaniEtAl2012_qlearn,GoldbergKosorok2012,simoneau2020estimating}.

Estimation of the terminal stage regression model is simply a typical model of outcome by predictors fit using standard Bayesian methods. The estimation of the nonterminal stage models, on the other hand, is not easily done with standard Bayesian software because of the counterfactual or potential payoff under the unobserved optimal action at each subsequent stage, which contributes to the outcome at the current stage. To address this problem, Murray et al.\  \cite{murray2018bayesian} developed a backward induction Gibbs (BIG) sampler to implement the proposed BML approach in practice. It consists of three steps, repeated until convergence, using $\hat{}$ above random variables to indicate sampled values in an MCMC algorithm:
\begin{enumerate}[label=Step \arabic*, leftmargin=*]
    \item Draw a posterior sample of parameters $\theta_2$ in the Stage 2 model and set the optimal action $\hat{a}_{i2}^{opt}=\hat{d}_2^{opt}(\bar{o}_{i2}; \theta_2)$, $i=1,\dots,n$.
    \item Compare the observed $a_{i2}$ and the optimal $\hat{a}_{i2}^{opt}$. For $i=1,\dots,n$, if $a_{i2}=\hat{a}_{i2}^{opt}$, then set $\hat{y}_{i2}^{opt}=y_{i2}$; else, sample $\hat{y}_{i2}^{opt}$ from the posterior predictive distribution of $y_2(a_{i1},\hat{a}_{i2}^{opt})$.
    \item Draw a posterior sample of parameters $\theta_1$ in the Stage 1 model using outcome $y_{i1}+\eta_i \hat{y}_{i2}^{opt}$.
\end{enumerate}

\subsection{AFT BART}
\par Bayesian additive regression trees (BART) form a Bayesian nonparametric regression model developed by Chipman et al.\   \cite{chipman2010bart}, which is an ensemble of trees. The accelerated failure time BART  \cite{bonato2011bayesian} is an extension of the approach to accommodate censored outcomes assuming the event time follows a log normal distribution. Let $t_i$ be the event time, $c_i$ be the censoring time for individual $i$. Then, the observed survival time is $ s_i =\min(t_i,c_i)$, and the event indicator is $\delta_i=I(t_i<c_i)$. Denote by $\bx_i=(x_{i1},\dots,x_{ip})$ the $p$-dimensional vector of predictors. The relationship between $t_i$ and $\bx_i$ is expressed as
\begin{align*}
    \log t_i &=  \mu+f(\bx_i)+\varepsilon_i,\quad &\varepsilon_i&\iid \cn(0,\sigma^2) \\
    f &\prior \BART, \quad &\sigma^2&\prior \nu\lambda\chi^{-2}(\nu),
\end{align*}
where $f(\bx_i)$ is a sum of $m$ regression trees $f(\bx_i)\equiv\sum_{j=1}^m g(\bx_i;\ct_j,\cm_j)$ with $\ct_j$ denoting a binary tree with a set of internal nodes and terminal nodes, and $\cm_j=\{\mu_{j1},\dots,\mu_{jb_j}\}$ denoting the set of parameter values on the terminal nodes of tree $\ct_j$. Full details of the BART model, including prior distributions and MCMC sampling algorithm, can be found in \cite{chipman2010bart}.  Since the $t_i$ of censored observations are not observable, an extra data augmentation step to impute $t_i$ is needed in each iteration when drawing Markov chain Monte Carlo (MCMC) posterior samples with Gibbs sampling. In particular, the unobserved event times are randomly sampled from a truncated normal distribution as
\[\log t_i|s_i,\delta_i=0,f(\bx_i),\sigma^2 \sim \cn(\mu+f(\bx_i),\sigma^2)\times I(t_i > s_i).\]
After data augmentation, the complete log event times are treated as continuous outcomes and the standard BART MCMC draws can be applied.

The AFT BART model with a log normal survival distribution is implemented within the BART R package \cite{BART}; additional details are found in the Appendix \ref{appen:func}.

\subsection{Proposed AFT-BML algorithm}
\label{sec:aftbml}
Since the BML approach by Murray et al.\   \cite{murray2018bayesian} is not directly applicable to censored observations, we extended it by modifying the BIG sampler so that censoring can be accommodated. Here we are interested in the time to an event (such as death) from the start of Stage 1.  The Stage 2 treatment decision initiates at an intermediate event such as disease progression.  This effectively separates the payoff or event time into two components: the time to the earliest of the event of interest and the intermediate event triggering Stage 2 ($t_1$), and if the patient enters Stage 2 ($\eta=1$), the time from the start of Stage 2 to the event of interest ($t_2$).  Observed data accounting for censoring and entry to Stage 2 are denoted $(s_{1},\delta_1)$ for Stage 1 and $(s_2,\delta_2)$ for Stage 2.  Continuing with the potential outcomes notation, let $t(a_1,a_2)$ denote the time to the event of interest when action $a_1$ is taken at Stage 1 and action $a_2$ is taken at Stage 2.  Similarly, let $t_2(a_1,a_2)$ denote the event time in Stage 2 (starting at the entry to Stage 2) under actions $(a_1,a_2)$.  Finally, potential time $t_1(a_1)$ is the time in Stage 1 until the first of the event of interest or entry to Stage 2.  Corresponding payoffs on the log time scale are denoted $y(a_1,a_2)=\log t(a_1,a_2)$, $y_2(a_1,a_2)=\log t_2(a_1,a_2)$, and $y_1(a_1)=\log t_1(a_1)$.  Under consistency, the observed outcome corresponds to the potential outcome for the action actually followed, e.g.  $t_1(a_1)=t_1$, $t_2(a_1,a_2)=t_2$, and $t(a_1,a_2)=t$, and similarly for the $y=\log t$ versions.

Murray et al.\   \cite{murray2018bayesian} recommended using Bayesian nonparametric regression models in Stages 1 and 2 for robustness.  Here we illustrated our approach with AFT BART models in each stage.  As before, we use $\hat{}$ above random variables to indicate sampled values in an MCMC algorithm.  The Stage 2 regression model for $t_2(a_1,a_2)$ is estimated first, using the observed covariates $(\bar{o}_2, a_2)$ and the observed time to event data $(s_2,\delta_2)$, according to the AFT BART model
\begin{align*}
    \log t_{i2} &=  \mu_2+f_2(\bar{o}_2,a_2)+\varepsilon_i,\quad &\varepsilon_i&\iid \cn(0,\sigma_2^2) \\
    f_2 &\prior \BART, \quad &\sigma_2^2&\prior \nu\lambda\chi^{-2}(\nu),
\end{align*}

We can run the Stage 2 BART model until convergence, draw 1000 posterior samples from the model, and then sample the optimal Stage 2 treatment rule for each MCMC sample according to
\begin{align*}
    \hat{d}_2^{opt}(\bar{o}_2) &= \arg \sup_{a_2\in\mathcal{A}_2} E(\log t_2|\bar{o}_2, a_2) = \arg \sup_{a_2\in\mathcal{A}_2} f_2(\bar{o}_2,a_2),
\end{align*}
We also can implement a sampling procedure to generate potential outcomes for the total time from Stage 1 assuming optimal Stage 2 treatment as $\hat{t}(a_1,\hat{d}_2^{opt})=t_1+\hat{t}_2(a_1,\hat{d}_2^{opt})$.  Some of the potential outcomes resulting from this procedure may still be censored, and we denote the possibly censored version of these potential outcomes as $(\hat{s},\hat{\delta})$.  This event time data are then modeled as a function of covariates $(o_1,a_1)$ using another AFT BART model given by
\begin{align*}
    \log \hat{t}_i &=  \mu_1+f_1(o_1,a_1)+\varepsilon_i,\quad &\varepsilon_i&\iid \cn(0,\sigma_1^2) \\
    f_1 &\prior \BART, \quad &\sigma_1^2&\prior \nu\lambda\chi^{-2}(\nu),
\end{align*}
For each sampled potential outcomes dataset, we run the Stage 1 AFT BART model until convergence, and then draw one posterior sample from each fitted BART model to determine a sample from the posterior of $d_1^{opt}$ according to
\begin{align*}
    \hat{d}_1^{opt}(o_1) &= \arg \sup_{a_1\in\mathcal{A}_1} E(\log \hat{t}|o_1,a_1, d_2^{opt}) = \arg \sup_{a_1\in\mathcal{A}_1} f_1(o_1,a_1).
\end{align*}
Details of the AFT-BML algorithm are as follows:

\begin{enumerate}[label=Step \arabic*, leftmargin=*]
    \item Run BART on the Stage 2 data until convergence and draw 1000 samples from the posterior distribution of $f_2$ and $\sigma_2^2$.  This implicitly involves the following two steps which are performed automatically by the BART package.
    \begin{enumerate}[label=Step 1\alph*]
    \item  Draw unobserved event time $t_{i2}$ for censored subjects ($\delta_{i2}=0$) who reached Stage 2 ($\eta_i=1$) using a truncated normal distribution,
    \[\log t_{i2} |s_{i2}, \delta_{i2}=0,f_2,\sigma_2^2 \sim \cn(\mu_2+f_2(\bar{o}_{i2},a_{i2}),\sigma_2^2) \times I(t_{i2}\ge s_{i2}).\]
    \item Update $f_2$ and $\sigma_2^2$ with complete (uncensored) Stage 2 data.
    \end{enumerate}
    \item Draw 1000 samples of  $(\hat{a}_{i2}^{opt},\hat{t}_{i2}^{opt})$ for each subject and use each sample to create an augmented dataset for Stage 1 analysis, as follows:
    \begin{enumerate}[label=Step 2\alph*]
        \item The optimal action at Stage 2 is chosen as $\hat{a}_{i2}^{opt}=\arg_{a_2}\max f_2(\bar{o}_{i2},a_2)$.
        \item If $a_{i2}=\hat{a}_{i2}^{opt}$ and the observation is an event ($\delta_{i2}=1$), $\hat{t}_{i2}^{opt}=t_{i2}$; if $a_{i2}=\hat{a}_{i2}^{opt}$ and the observation is censored ($\delta_{i2}=0$), $\log \hat{t}_{i2}^{opt} \sim \cn(\mu_2+f_2(\bar{o}_{i2},\hat{a}_{i2}^{opt}),\sigma_2^2) \times I(\hat{t}_{i2}^{opt}\ge s_{i2})$.
        \item If $a_{i2}\ne \hat{a}_{i2}^{opt}$, draw $\log \hat{t}_{i2}^{opt}$ for the counterfactual action $\hat{a}_{i2}^{opt}$ from $\cn(\mu_2+f_2(\bar{o}_{i2},\hat{a}_{i2}^{opt}),\sigma_2^2)$.
        \item For those who reached Stage 2 ($\eta_i=1$), set the observed data for the Stage 1 model as the potential Stage 1 event time $\hat{t}_i$, e.g. $\hat{s}_i=\hat{t}_i=t_{i1}+\hat{t}_{i2}^{opt}$, and set $\hat{\delta}_i=1$.  For those who did not reach Stage 2, set the observed data for the Stage 1 model as $\hat{s}_i=t_{i1}$ and $\hat{\delta}_i=\delta_{i1}$.
        \end{enumerate}
        \item  Run BART on each of the augmented Stage 1 data sets until convergence and draw 1 sample from the posterior distribution of $f_1$ and $\sigma_1^2$ for each augmented Stage 1 data set.  As above, this requires the following two steps which are performed automatically by the BART package:
        \begin{enumerate}[label=Step 3\alph*]
        \item Draw unobserved event time $\hat{t}_{i}$ for censored subjects ($\hat{\delta}_{i}=0$) at Stage 1,
        \[\log \hat{t}_{i} | \hat{s}_i,\hat{\delta}_i=0,f_1,\sigma_1^2 \sim \cn(\mu_1+f_1(o_{i1},a_{i1}),\sigma_1^2) \times I(\hat{t}_{i}\ge \hat{s}_{i}).\]
        \item \label{b} Update $f_1$ and $\sigma_1^2$ with complete (uncensored) data at Stage 1.
\end{enumerate}
        \item From each augmented dataset and the corresponding sampled $f_1$ and $\sigma_1^2$, draw one sample $\hat{a}_{i1}^{opt}$ for each subject, based on \[\hat{a}_{i1}^{opt}=\arg\max_{a_{i1}} f_1(o_{i1},a_{i1}).\]
\end{enumerate}

The original BIG sampler indicated that the sampling of the Stage 1 parameters should be updated using the values from the prior iteration.  However, while that could potentially speed up implementation as it may not require a full burn-in for each new Stage 1 dataset, it is challenging to implement because most BART software does not allow for starting an update step from a specified value of the tree structure and the terminal node means. Instead, we leverage the fact that the BART chain for Stage 2 does not depend on any updates of the Stage 1 model parameters. Because of this, the BART model for Stage 2 can be run independently and used to generate the potential datasets for Stage 1.  Once the 1000 datasets for Stage 1 have been sampled, the BART analyses of each of these Stage 1 datasets in Step 3 can be done in parallel using off the shelf BART software.

Our approach for drawing event times for patients who were censored and who received optimal treatment in Stage 2 was to first sample exact event times for Stage 2 data from the Stage 2 model and then pass this value as an event to the Stage 1 dataset (after adding the observed time in the first stage).  Alternatively, one could pass the value as censored to the Stage 1 dataset, in which case the AFT BART model would implicitly sample event times using the Stage 1 model, instead of using the Stage 2 model as in the algorithm above.  We also implemented and examined this alternative approach in our simulation studies, but found no measurable difference in the results, so we did not consider it further.

We implemented the proposed method by creating a wrapper function called \texttt{dtr1} that utilizes the BART R package \cite{BART}.  Details on the software implementation can be found in the Appendix.

\section{Simulations}
\label{sec:simu}

\subsection{Simulation Design}
We conducted simulation studies with 200 replicated training sets of sample size $N=800$ and an independent testing set of sample size $n=400$ for each scenario of interest to demonstrate the predictive performance of our method. An observational study with two stages of treatment setting was used with two candidate treatments at each stage. The treatment assignments were generated from a Bernoulli distribution with a probability $P(A_1=1)$ and $P(A_2=1)$, respectively. Both the event time at Stage 2, as well as the overall event time assuming optimal treatment at Stage 2, were generated from AFT models, assuming a log-normal distribution, similar to the approach of Simoneau et al.\   \cite{simoneau2020estimating}. The censoring time was assumed to follow a Uniform distribution that led to approximately 20\% to 30\% independent censoring.

We fit each training data set with our method, and made predictions of the optimal action and the mean of the log-normal event time distribution under optimal treatment at each stage on the test data set. Our performance was compared against Q-learning, including an oracle model along with other models that misspecified the relationship  for either stage. We looked at the proportion of optimal treatment (POT), mean-squared error (MSE), and 95\% credible intervals coverage rate (CR) (for the BART only approach). The POT is defined at each stage as
\begin{equation}\label{pot}
    POT_j=\frac{1}{n}\sum_{i=1}^n I\{\hat{a}_{ij}^{opt}=a_{ij}^{opt}\},\quad j=1,2.
\end{equation}

\subsection{Simulation Settings}
For subject $i$, a continuous baseline covariate $x_{i1}$ was drawn from a $\mathrm{U}(0.1,1.29)$ distribution, and a binary baseline covariate $b_{i1}$ was from a Bernoulli distribution with probability $0.5$. Similarly, a continuous covariate $x_{i2}$ that was measured at the beginning of Stage 2 was also generated from a $\mathrm{U}(0.9,2)$ distribution, and a binary covariate $b_{i2}$ measured at the beginning of Stage 2 was randomly drawn from a $\mathrm{Bern}(0.5)$ distribution. Additionally there were two noise covariates, $z_{i1}\sim \cn(10,3^2),z_{i2}\sim \cn(20,4^2)$, collected at the beginning of Stage 1 and Stage 2, respectively. When fitting the data, all the stage-wise covariates were included in the models to mimic real-world settings in which there is uncertainty as to which covariates are relevant predictors of the outcomes. The Stage 1 treatment was assigned from a Bernoulli distribution with the probability of receiving treatment $P(a_{i1}=1)=\expit(2x_{i1}-1)$, where $\expit(x)=\exp(x)/(1+\exp(x))$ is the inverse of the logit function. For those who entered the second stage ($\eta_i=1$), the Stage 2 treatment was sampled from a Bernoulli distribution with $P(a_{i2}=1)=\expit(-2x_{i2}+2.8)$. The probability of entering Stage 2 was fixed at $0.6$, i.e., $P(\eta_i=1)=0.6$.

We considered two different scenarios for the relationship between the log event time and the covariates.  In Scenario 1, we used an AFT model to generate the event time at Stage 2 as
\begin{align}
    \log t_{i2} &= 4+0.3x_{i2}+b_{i2}-0.6x_{i2}b_{i2}+0.3x_{i1}+0.4b_{i1}-0.5x_{i1}b_{i1} \notag\\
    & \quad +a_{i2}(-0.7+0.5x_{i2}-0.9b_{i2})+\epsilon_{i2}, \quad \epsilon_{i2} \sim \cn(0, 0.3^2). \label{sim1_1}
\end{align}
The true optimal treatment $a_{i2}^{opt}$, given by $I(-0.7+0.5x_{i2}-0.9b_{i2}>0)$, was plugged into equation (\ref{sim1_1}) as a new $a_{i2}$ to calculate the optimal Stage 2 event time $\hat{t}_{i2}^{opt}$ had everyone received their optimal treatment at Stage 2. The overall event time assuming optimal Stage 2 treatment was generated again from an AFT model as
\begin{align}
    \log \hat{t}_{i} &= 6.3+0.7x_{i1}+0.6b_{i1}-0.8x_{i1}b_{i1} \notag\\
    & \quad +a_{i1}(0.1-0.2x_{i1}+0.6b_{i1}) + \epsilon_{i1}, \quad \epsilon_{i1} \sim \cn(0, 0.3^2). \label{sim1_2}
\end{align}
For those who did not enter Stage 2, $\hat{t}_{i}$ was their event time. For those who entered Stage 2, the observed Stage 1 survival time was $t_{i1}=\hat{t}_{i} - \hat{t}_{i2}^{opt}$, and the Stage 2 event time was $t_{i2}$. The censoring time $c_i$ was generated from $U(100,2000)$ to yield an overall censoring rate of around 20\%.

As a comparator, we fitted the data with parametric Q-learning models as well. Since there were two stages in our simulation data, we chose either correctly specified (T) or misspecified (F) Q-function models for each stage as
\begin{enumerate}[label=Stage \arabic*, leftmargin=*]
    \item Q$_{1T}$: $x_{i1} + b_{i1} + x_{i1}b_{i1} + a_{i1} + a_{i1}x_{i1} + a_{i1}b_{i1}$ \\
    Q$_{1F}$: $x_{i1} + b_{i1} + z_{i1} + a_{i1} + a_{i1}x_{i1} + a_{i1}z_{i1}$
    \item Q$_{2T}$: $x_{i2} + b_{i2} + x_{i2}b_{i2} + x_{i1} + b_{i1} + x_{i1}b_{i1} + a_{i2} + a_{i2}x_{i2} + a_{i2}b_{i2}$ \\
    Q$_{2F}$: $x_{i2} + b_{i2} + z_{i2} + x_{i1} + b_{i1} + a_{i2} + a_{i2}x_{i2} + a_{i2}z_{i2}$
\end{enumerate}
Combining the two stages together yields four possible modelling specifications: Q$_{1T2T}$, Q$_{1T2F}$, Q$_{1F2T}$, and Q$_{1F2F}$. Among these four Q-learning models, Q$_{1T2T}$ correctly specifies the parametric form in both stages; we refer to this as the oracle model.

In Scenario 2, we followed a similar structure to simulate the data but with a different set of true models that include non-linear transformations of the covariates.  The event time at Stage 2 was generated based on the following equation as
\begin{align}
    \log t_{i2} &= 4+\cos(x_{i2}^3)-0.4(x_{i2}b_{i2}+0.5)^2-0.1x_{i1}-\sin(\pi x_{i1}b_{i1}) \notag \\
    & \quad +a_{i2}(0.7x_{i2}^2-1)+\epsilon_{i2}, \quad \epsilon_{i2} \sim \cn(0, 0.1^2). \label{sim2_1}
\end{align}
The true optimal treatment, $a_{i2}^{opt}=I(0.7x_{i2}^2-1>0)$, was used to replace $a_{i2}$ in equation (\ref{sim2_1}) to calculate the optimal Stage 2 event time $\hat{t}_{i2}^{opt}$. The overall event time $\hat{t}_{i}$ assuming optimal Stage 2 treatment was generated as
\begin{align}
    \log \hat{t}_{i} &= 7.4+\sin(x_{i1}^2)+x_{i1}^4+x_{i1}b_{i1} \notag \\
    & \quad +a_{i1}(0.1-0.2x_{i1}^3) + \epsilon_{i1}, \quad \epsilon_{i1} \sim \cn(0, 0.1^2). \label{sim2_2}
\end{align}
The Stage 1 and Stage 2 survival times were calculated in the same way as in Scenario 1, depending on whether the observation entered the second stage. Censoring time $c_i$ was now generated from $U(400,5000)$ to achieve an overall censoring rate of around 30\%.

Based on the underlying true nonlinear functions of covariates, we constructed two misspecified Q-learning models besides the oracle model. The first misspecified model Q$_{lin}$ considered only linear terms in the covariates for both stages as
\begin{enumerate}[label=Stage \arabic*, leftmargin=*]
    \item Q$_{lin}$: $x_{i1} + b_{i1} + z_{i1} + a_{i1}$,
    \item Q$_{lin}$: $x_{i2} + b_{i2} + z_{i2} + x_{i1} + b_{i1} + z_{i1} + a_{i2}$.
\end{enumerate}
The second misspecified model Q$_{int}$ considered all two-way interactions among covariates and all interactions between treatment and covariates in each stage in addition to the linear terms in Q$_{lin}$ as
\begin{enumerate}[label=Stage \arabic*, leftmargin=*]
    \item Q$_{int}$: $x_{i1} + b_{i1} + z_{i1} + x_{i1}b_{i1} + x_{i1}z_{i1} + b_{i1}z_{i1} + a_{i1} + a_{i1}x_{i1} + a_{i1}b_{i1} + a_{i1}z_{i1}$,
    \item Q$_{int}$: $x_{i2} + b_{i2} + z_{i2} + x_{i2}b_{i2} + x_{i2}z_{i2} + b_{i2}z_{i2} + x_{i1} + b_{i1} + z_{i1} + x_{i1}b_{i1} + x_{i1}z_{i1} + b_{i1}z_{i1} + a_{i2} + a_{i2}x_{i2} + a_{i2}b_{i2} + a_{i2}z_{i2}$,
\end{enumerate}
such that these models were not correctly specified but were nonetheless richer and more flexible than their only `linear' counterparts.

\subsection{Simulation Results}
For the proposed method, denoted as BART in the figures, we created a wrapper function \texttt{dtr1} that implemented the algorithm described in Section \ref{sec:aftbml}; further documentation of this implementation is available in the Appendix. For Q-learning, we first used \texttt{survreg} function from the R package \texttt{survival} \cite{survival-book,survival-package} to fit the Stage 2 model, then made predictions of the optimal second stage treatment and corresponding optimal survival time to create Stage 1 data. The \texttt{survreg} function was called again to fit the new augmented Stage 1 data and estimate the optimal first stage treatment with corresponding optimal overall survival time. The general framework is the same as the proposed method. Given that all the covariates at both stages were simulated for every subject, the related Stage 2 treatment and time were predicted for everyone in the test set. This is unrealistic in practice since some covariates are not available if a patient never entered Stage 2.

The proportion of optimal treatment is defined as the ratio of the number of subjects that has the true optimal treatment correctly identified by the model in a specific stage and the total number of subjects in the test set, that is $400$. For the stage-wise POT, the subject is counted in the numerator if the optimal treatment matches with the truth in a specific stage, as shown in equation (\ref{pot}). For the overall or combined POT, only those who have the true optimal treatment correctly identified at both stages are included in the numerator, as in the following expression:
\begin{equation*}
    \frac{1}{n}\sum_{i=1}^n I\{\hat{a}_{i1}^{opt}=a_{i1}^{opt}\}I\{\hat{a}_{i2}^{opt}=a_{i2}^{opt}\}.
\end{equation*}
It is straightforward to calculate POTs with Q-learning since that approach makes only one prediction of the optimal treatment at each stage for each observation. With our Bayesian approach, a little additional effort is needed. 
Since BART draws posterior samples to describe the posterior distribution, the raw results from our approach are also a set of posterior samples. Specifically, every single posterior sample consists of the outcomes under each candidate treatment at each stage for each observation. Comparing the corresponding outcomes across the candidate treatments, a collection of the optimal treatment and optimal outcome at each stage can be obtained for each observation and each posterior sample. With a set of posterior samples of optimal treatment, we can compute the posterior mean of optimal treatment probability, and the one with the highest posterior mean is the final prediction of the optimal treatment at each stage for each observation. That is the quantity we used to calculate the POTs for AFT-BML.
The mean squared error (MSE) was calculated by comparing the estimated optimal Stage 2 and overall log event time means to the true means.

\begin{figure}[h]
    \centering
    \includegraphics[width=\textwidth]{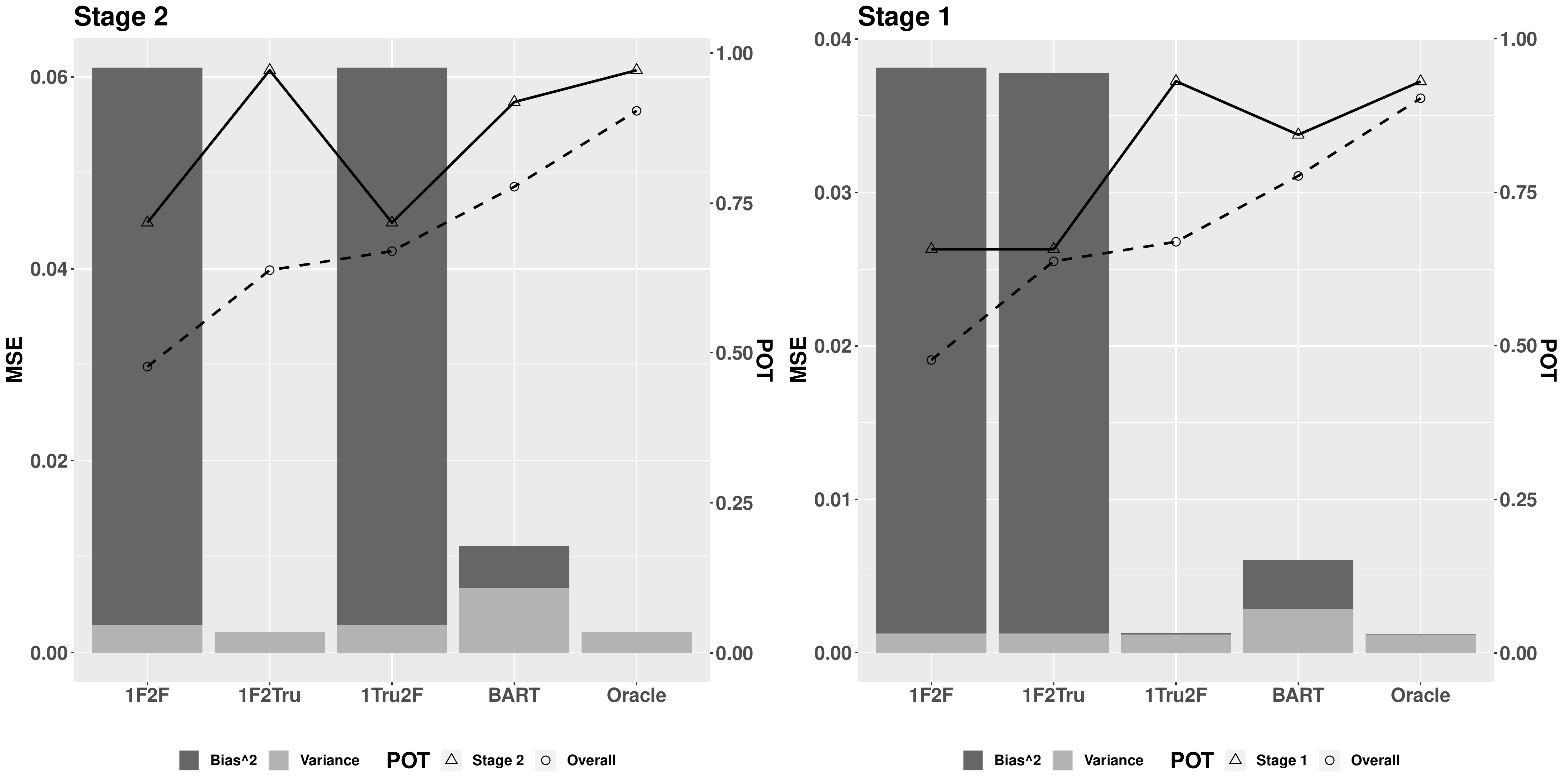}
    \caption{Mean squared error decomposed into variance and bias$^2$ for Scenario 1, in which there is linear dependence of the outcome on covariates at Stage 1 and 2. The left y-axis corresponds to MSE (vertical bars); the secondary y-axis corresponds to stage-wise and overall POT (solid and dashed lines, respectively).}
    \label{fig:lin}
\end{figure}
Figure \ref{fig:lin} shows the decomposition of MSE at both stages, as well as the stage-wise POT and overall POT, for the proposed method, oracle Q-learning model, and the three other misspecified models in Scenario 1.
Notice that in both equations (\ref{sim1_1}) and (\ref{sim1_2}), the relationship between log event, time and covariates are linear. As a parametric method that has the right structure as the underlying true model, the performance of the oracle model bests the others, with a very small MSE with zero bias, close to 100\% stage-wise POT at both stages and a 92\% overall POT. For Stage 2, Q$_{1F2T}$ has the same MSE and stage-wise POT as the oracle model since they specified the functional form in the exact same way. Among the other three models, our method performs the best, in terms of a smaller MSE with an even smaller bias, and a higher stage-wise POT with the difference greater than 20\%. For Stage 1, the MSE from Q$_{1T2F}$ is slightly bigger but very similar to the oracle, and the stage-wise POT is almost the same as the oracle model, even though the predicted Stage 2 optimal survival time from Q$_{1T2F}$ was based on a misspecified Stage 2 model. This is mainly due to the fact that the simulated Stage 2 event time was relatively small compared to the overall event time so that an incorrect prediction for Stage 2 has a minimal impact on the augmented overall survival time. This resulted a very similar data set between Q$_{1T2F}$ and the oracle when fitting the Stage 1 model. The proposed Bayesian method, as in Stage 2, has the smallest MSE and the highest stage-wise POT compared to the other two models (Q$_{1F2F}$ and Q$_{1F2T}$). Our proposed method is better than all approaches except the oracle approach when taking optimal treatment for both stages into consideration using the overall POT.

\begin{figure}[h]
    \centering
    \includegraphics[width=\textwidth]{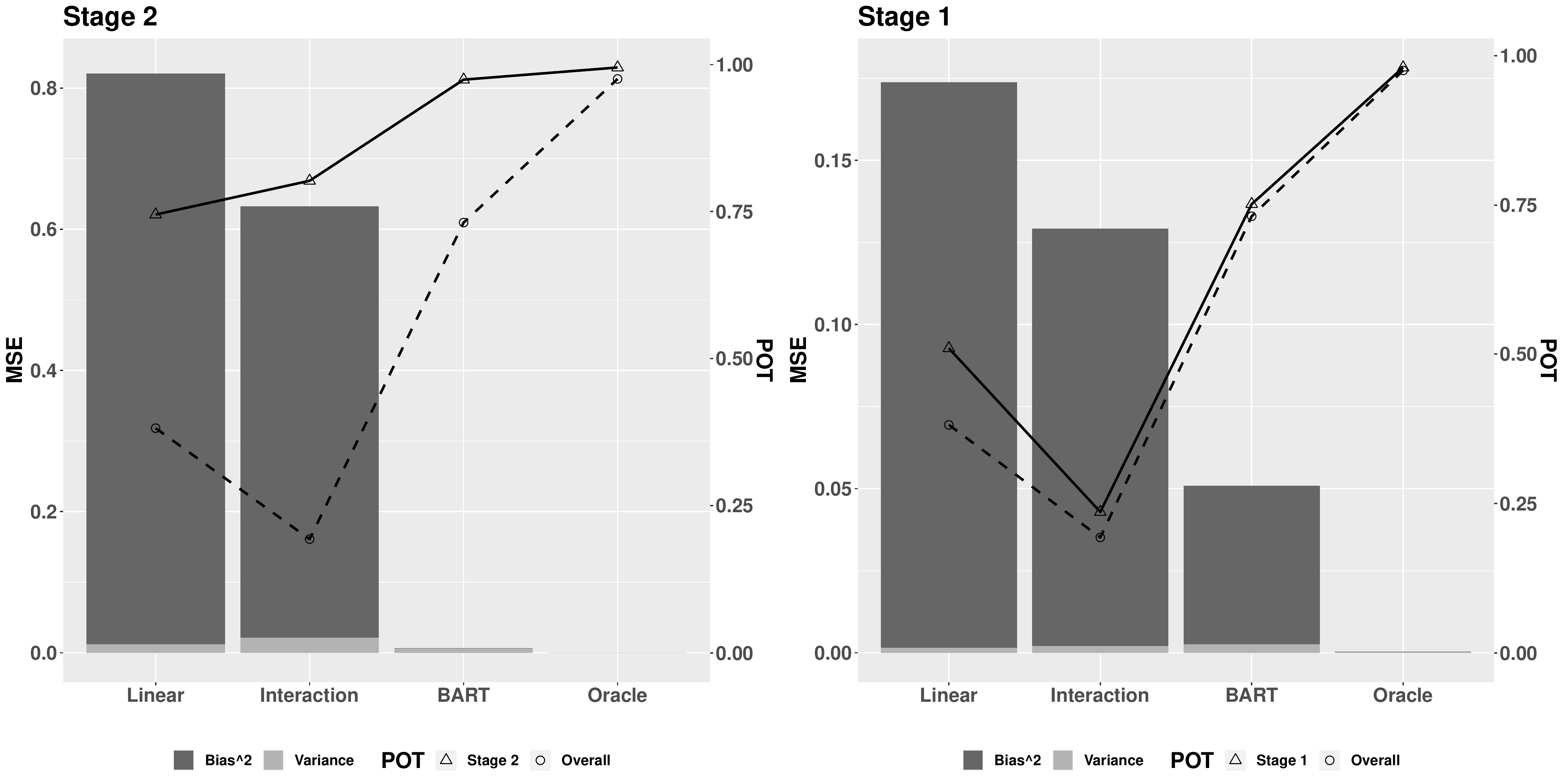}
    \caption{Mean squared error decomposed into variance and bias$^2$ for Scenario 2, in which there is  nonlinear dependence of the outcome on covariates at Stage 1 and 2. The left y-axis corresponds to MSE (vertical bars); the secondary y-axis corresponds to stage-wise and overall POT (solid and dashed lines, respectively).}
    \label{fig:nln}
\end{figure}

The results from Scenario 2 are shown in figure \ref{fig:nln}. The relationship is nonlinear between the covariates and log event time in both equations (\ref{sim2_1}) and (\ref{sim2_2}). As expected, the oracle model has close to zero MSE and close to 100\% POTs. Our method outperforms the other two models (Q$_{lin}$ and Q$_{int}$) with a much smaller MSE and higher POTs. The magnitude of the differences in figure \ref{fig:nln} are larger than those in figure \ref{fig:lin}. The advantage of our nonparametric method becomes more obvious in exploring the nonlinear dependencies, while the other two parametric Q-learning models suffered from incorrect model structures.

\begin{figure}[h]
    \centering
    \includegraphics[width=0.45\textwidth]{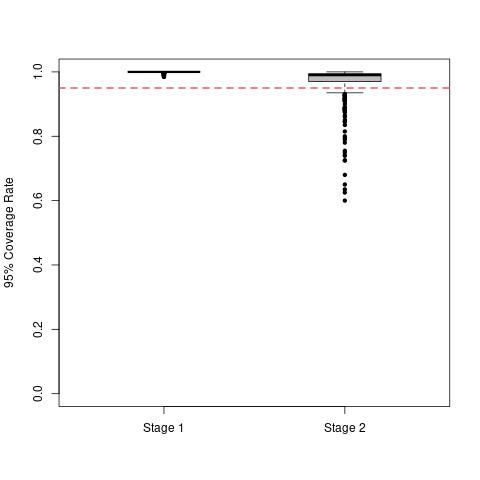}
    \includegraphics[width=0.45\textwidth]{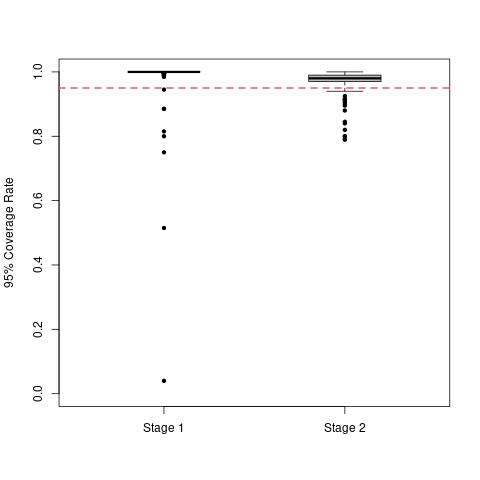}
    \caption{The coverage rate (CR) of 95\% credible intervals for the log event time mean under optimal treatment from the proposed method for Scenario 1 (Left) and Scenario 2 (Right) with the red reference line indicating the nominal 95\% level.}
    \label{fig:cr}
\end{figure}

Another quantity that is not easily estimable for Q-learning but comes without extra cost for BART is the measure of uncertainty in the estimated optimal values. By drawing MCMC posterior samples, the standard error of log optimal survival times can be calculated as the sample standard deviation. The credible intervals of log event time are also derivable with a collection of posterior samples. On the contrary, to obtain the standard error and confidence interval with Q-learning, bootstrap sampling must be carried out. Here, we only show the coverage rate (CR) of 95\% credible intervals for our method in figure \ref{fig:cr}. Q-learning models are not presented because of the poor fit and high biases in figure \ref{fig:lin} and \ref{fig:nln}. The boxplot of 95\% CR for both scenarios are almost always above the nominal 95\% at Stage 1, and always cover with the lower quartiles above the nominal 95\% at Stage 2. This indicates that the proposed method has good accuracy in estimating the uncertainty in the log event time mean under optimal treatment for both stages although the CRs are slightly over 95\%.

\section{Motivating Analysis: Optimal Treatment for AML Patients undergoing Transplant}
\label{sec:ex}
In this section, we applied the proposed method to data  from the Center for International Blood and Marrow Transplant Research (CIBMTR) (Krakow et al.  \cite{krakow2017tools}). There are 4171 patients with complete information in this data who received graft-versus-host disease prophylaxis for their allogeneic hematopoietic cell transplant, which was used to treat their myeloid leukemia, between 1995 and 2007. Some patients were subsequently given a salvage treatment after they developed GVHD and experienced unsuccessful initial treatment. The two stages considered in this study were up-front GVHD prophylaxis treatment and salvage treatment after developing GVHD and failing initial treatment (which is consistently given as steroids). In each stage, patients were assigned one of two treatments, nonspecific highly T-cell lymphodepleting (NHTL) immunosuppressant therapy or standard prophylaxis immunosuppressant. Estimating an optimal DTR to maximize the overall disease-free survival time for patients is our primary goal. The primary outcome is time to death, disease persistence, or relapse. The Stage 1 time is defined as the time from graft infusion to diagnosis of steroid-refractory acute GVHD (if they enter Stage 2) or to the primary outcome or last follow-up (if they do not enter Stage 2).  The Stage 2 time is defined as the time from starting salvage treatment for steroid refractory GVHD to the primary outcome or last follow up. Among the 13 covariates of interest, time from graft infusion to acute GVHD onset ($\ge 1$ month, $<1$ month) and use of $\ge 4$ immunosuppressors to treat acute GVHD on index form (Yes, No) are only available for those patients who failed at the first treatment. The other covariates include recipient's age group ($<10$ years, $10-39$ years, $\ge 40$ years), Karnofsky/Lansky performance status at time of transplant ($\ge 80\%$, $<80\%$), disease status at time of transplant (Early, Intermediate, Advanced), donor relationship (Related, Unrelated), donor-recipient sex (female-male, other), graft source (Bone marrow, Peripheral blood, Umbilical cord), human leukocyte antigen (HLA) match (Well-matched, Partially matched, Mismatched), total-body irradiation (Yes, No), cytomegalovirus status (Negative-negative, Donor or recipient positive), conditioning intensity (Myeloblative, RIC/nonmyeloablative), and use of corticosteroids as part of GVHD prophylaxis (No, Yes). The prophylaxis assigned in Stage 1 is also used in fitting the salvage Stage 2 model. A frequency cross table of prophylaxis and salvage treatment assigned is shown in table \ref{tab:freq}.
\begin{table}[ht]
    \centering
        \caption{Treatment assigned at Stage 1 and 2}
        \begin{tabular}{cc|cc|c}
     & & \multicolumn{2}{c|}{Stage 2} & Not entered \\
         & & Standard & NHTL & Stage 2\\
         \hline
     \multirow{2}{3.5em}{Stage 1} &  Standard & 673 & 219 & 2180\\
     &   NHTL & 240 & 91 & 768
    \end{tabular}

    \label{tab:freq}
\end{table}

Both Q-learning and AFT-BML approaches were used to fit this two stage survival data DTR estimation.  All the main effects and the two-way interactions between stage-wise treatment and the other covariates are included in Q-learning models, and 1000 nonparametric bootstrap resamples were generated to estimate the uncertainty of the quantities of interest. For nonparametric AFT-BML, with 1000 MCMC posterior samples, the full distribution was available for any predictions. The point estimates of parameters along with bootstrap mean and 95\% confidence interval (CI) at each stage were examined for Q-learning. The waterfall plots for the mean differences in DFS on the log time scale under each treatment at each stage for each individual were created for both Q-learning and AFT-BML, as well as the 95\% and 50\% credible intervals (bootstrap CIs for Q-learning) presented on the same plot. The differences in the median DFS were also explored for both methods, as were the differences in the two year DFS probabilities.

The analysis results from Q-learning, including point estimates and bootstrap mean, as well as the bootstrap 95\% CI, are shown in the appendix tables \ref{tab:qstg1} and \ref{tab:qstg2}. Inspection of the 95\% CI for the interaction terms in table \ref{tab:qstg1} reveals covariate combinations that can be used to identify subgroups where the NHTL or the standard treatment is preferable. For example, assuming all the other covariates are at the reference level, an unrelated donor would benefit more from NHTL than the standard treatment at Stage 1. Similarly in table \ref{tab:qstg2}, when holding the other covariates at the reference level, a patient who received NHTL at Stage 1 would be expected to have a longer DFS time if the standard treatment was given at Stage 2, since the 95\% CI of A2.NHTL*A1.NHTL is negative.  Intuitively, this might be the case because salvage treatment that is different than the initial treatment which already failed might be expected to be more effective.

\begin{figure*}[h]
    \centering
    \begin{subfigure}[b]{0.475\textwidth}
        \centering
        \includegraphics[width=\textwidth]{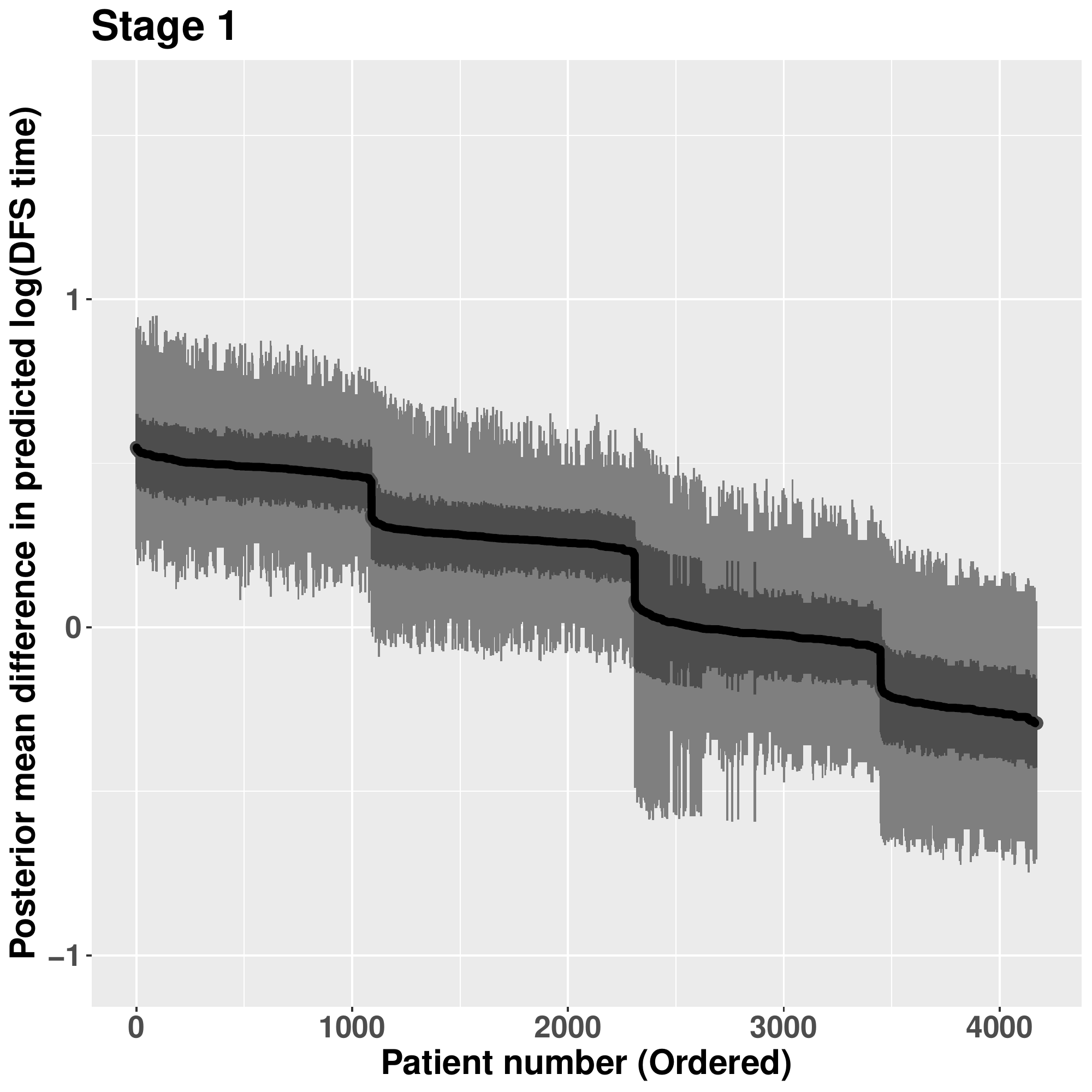}
        \caption[Network2]%
        {{\small AFT-BML (NHTL - Standard)}}
        \label{fig:bart1dfslg}
    \end{subfigure}
    \hfill
    \begin{subfigure}[b]{0.475\textwidth}
        \centering
        \includegraphics[width=\textwidth]{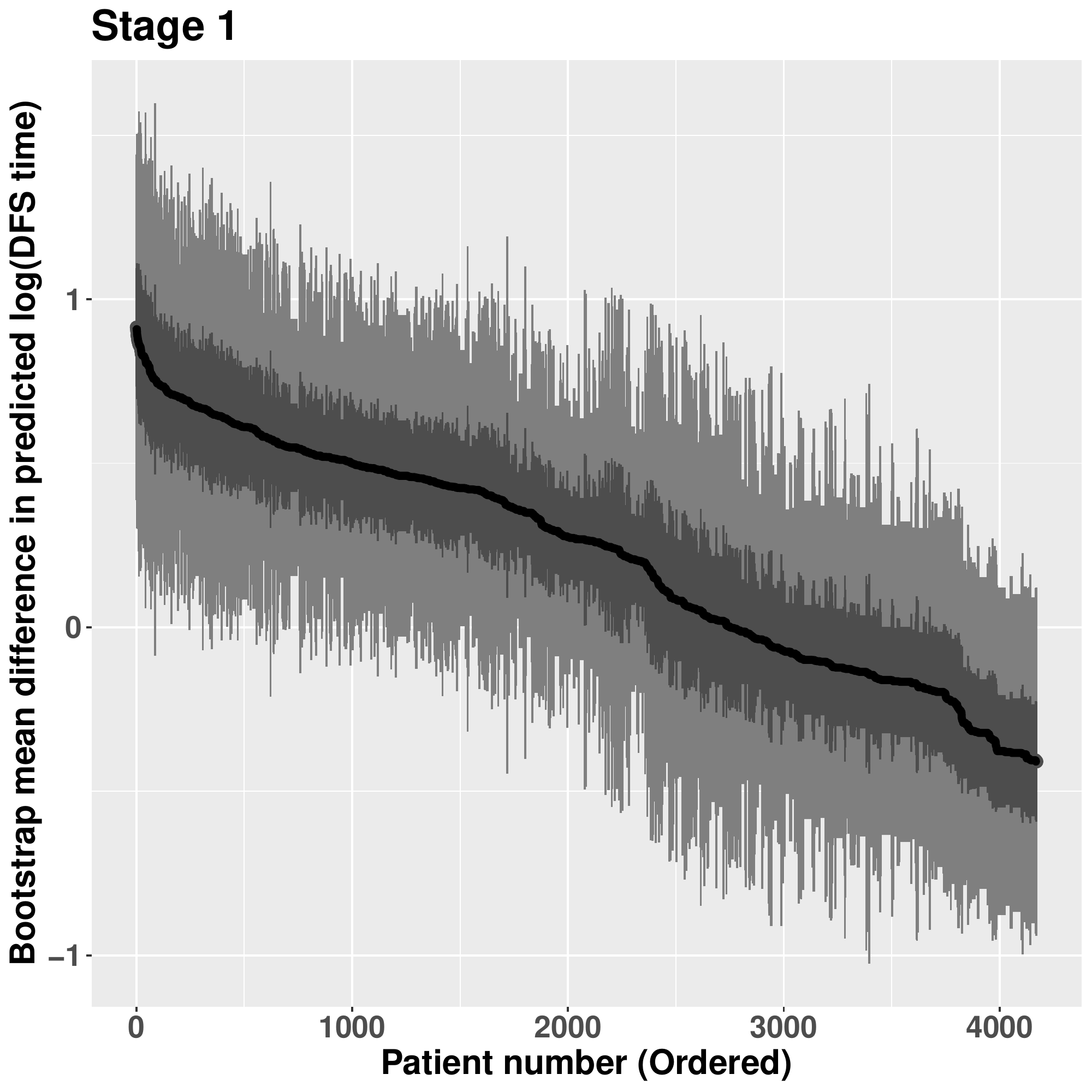}
        \caption[]%
        {{\small Q-learning (NHTL - Standard)}}
        \label{fig:qlearn1dfslg}
    \end{subfigure}
    \vskip\baselineskip
    \begin{subfigure}[b]{0.475\textwidth}
        \centering
        \includegraphics[width=\textwidth]{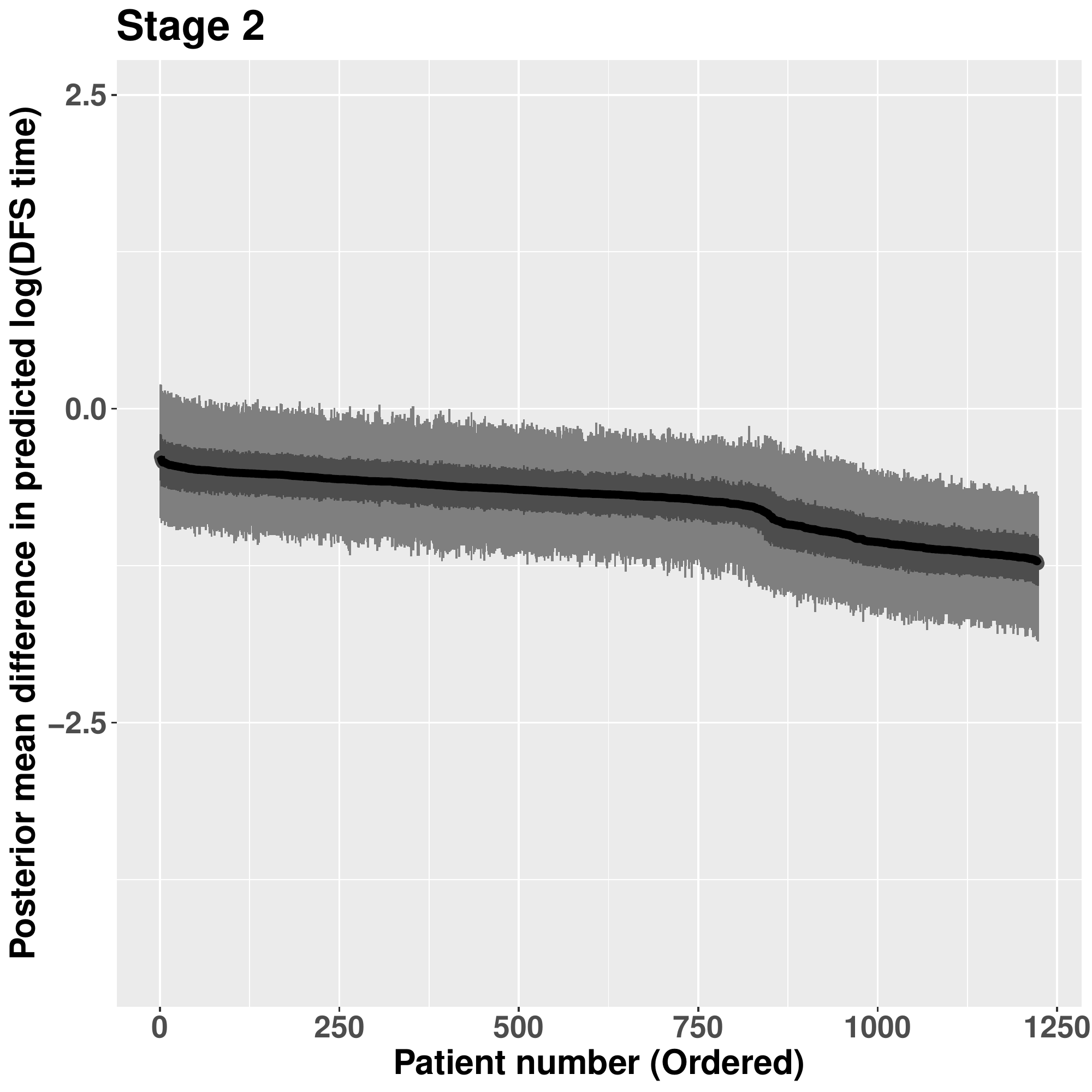}
        \caption[]%
        {{\small AFT-BML (NHTL - Standard)}}
        \label{fig:bart2dfslg}
    \end{subfigure}
    \hfill
    \begin{subfigure}[b]{0.475\textwidth}
        \centering
        \includegraphics[width=\textwidth]{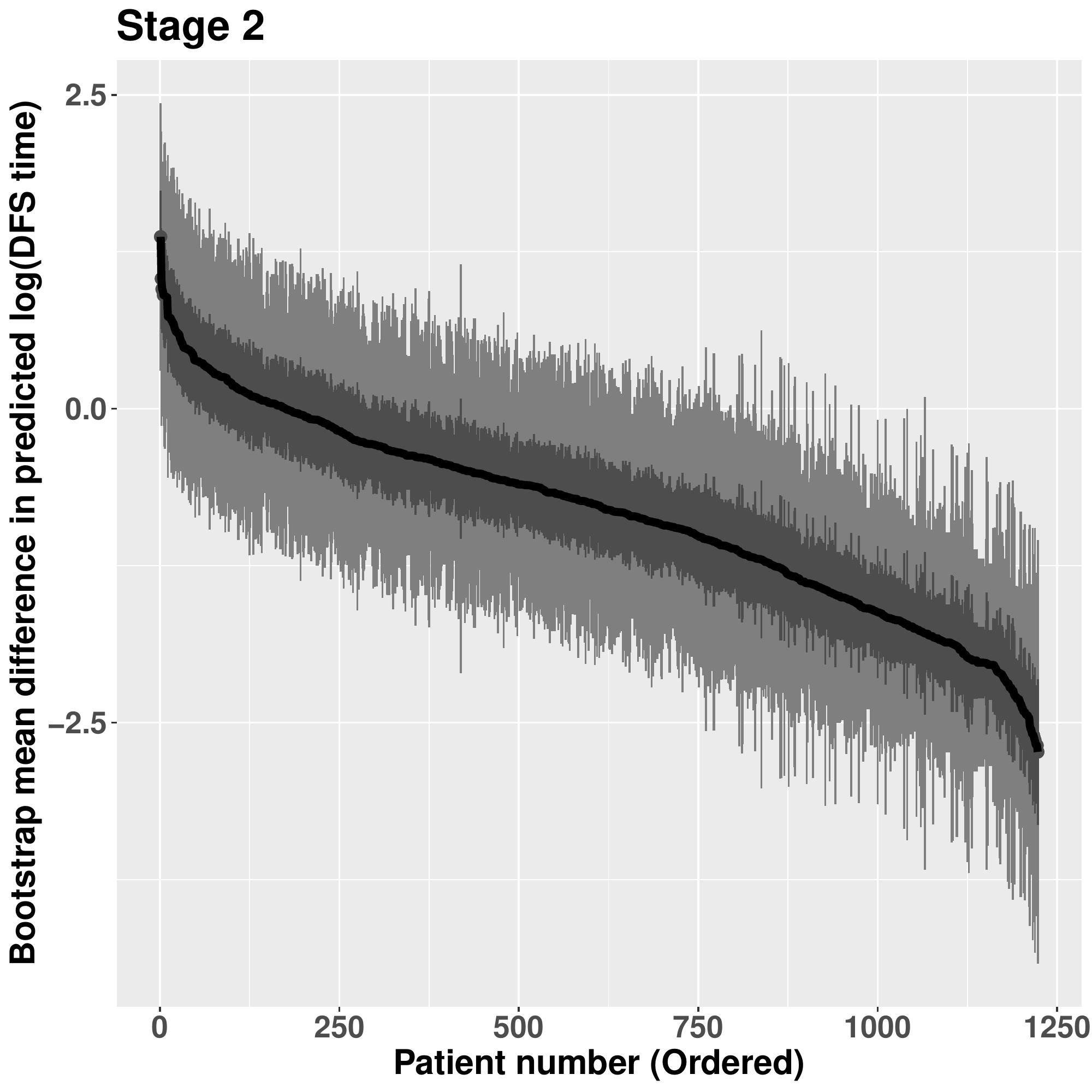}
        \caption[]%
        {{\small Q-learning (NHTL - Standard)}}
        \label{fig:qlearn2dfslg}
    \end{subfigure}
    \caption[ ]
    {\small Predicted mean difference of log DFS time among patients who received AHCT with NHTL versus standard treatment.}
    \label{fig:dfslg}
\end{figure*}

In figure \ref{fig:dfslg}, we present the estimated treatment differences on the log time scale for each stage, along with 95\% CI and 50\% CI.  For AFT-BML, a direct posterior prediction difference for each individual can be calculated from 1000 MCMC posterior samples based on patient-level characteristics. The credible intervals are constructed using the quantiles of posterior samples of each patient. For Q-learning, a standard error can be estimated from the predictions of 1000 bootstrap resamples at an individual level. Using the estimated standard error, the CIs for each patient are calculated in a standard way. A positive difference means NHTL is the preferred treatment for a given stage. The patients are presented in descending order based on the estimated difference, separately for each method.

The results from AFT-BML (figures \ref{fig:bart1dfslg} and \ref{fig:bart2dfslg}) suggest that there is little to be gained by individualizing the treatment in Stage 2 since everyone benefits from the standard treatment, but at Stage 1 there may be significant clinical value in choosing treatment in a personalized fashion to maximize the overall DFS time. In fact, there may be four subgroups that have a distinct difference in expected log event time, in which two groups would benefit from NHTL, one group is indifferent to treatment choice, and one final group that would have longer survival time with the standard treatment. The right panels, showing figures \ref{fig:qlearn1dfslg} and \ref{fig:qlearn2dfslg}, provides a similar message using Q-learning: while the standard treatment is the preferred treatment for most (though not all) patients at Stage 2, individualizing the treatment at Stage 1 may lead to important benefits in log DFS time.  However, there is a more continuous spectrum of treatment differences in Stage 1 using Q-learning, compared to AFT-BML.

\begin{figure*}[h]
    \centering
    \begin{subfigure}[b]{0.475\textwidth}
        \centering
        \includegraphics[width=\textwidth]{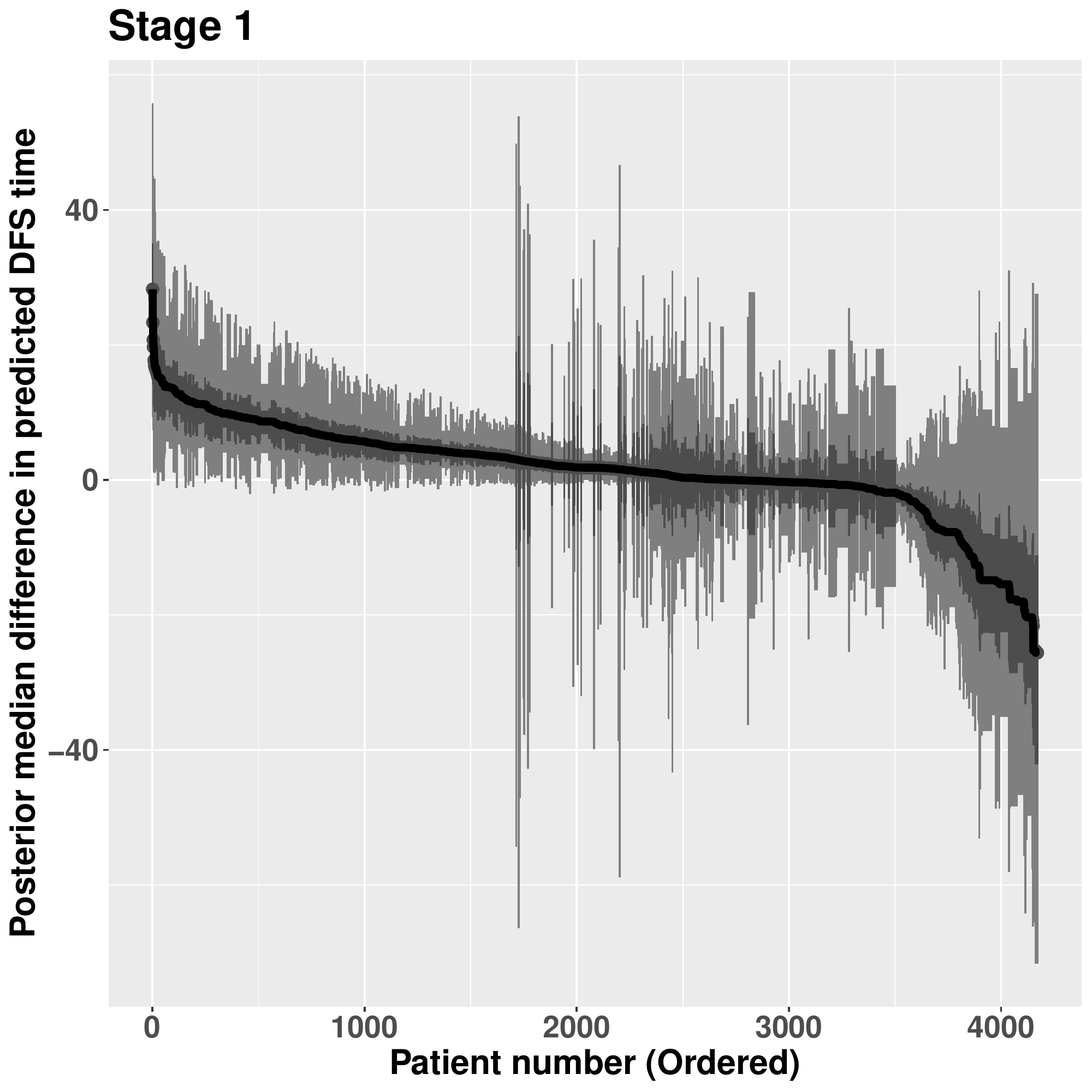}
        \caption[Network2]%
        {{\small AFT-BML (NHTL - Standard)}}
        \label{fig:bart1dfs}
    \end{subfigure}
    \hfill
    \begin{subfigure}[b]{0.475\textwidth}
        \centering
        \includegraphics[width=\textwidth]{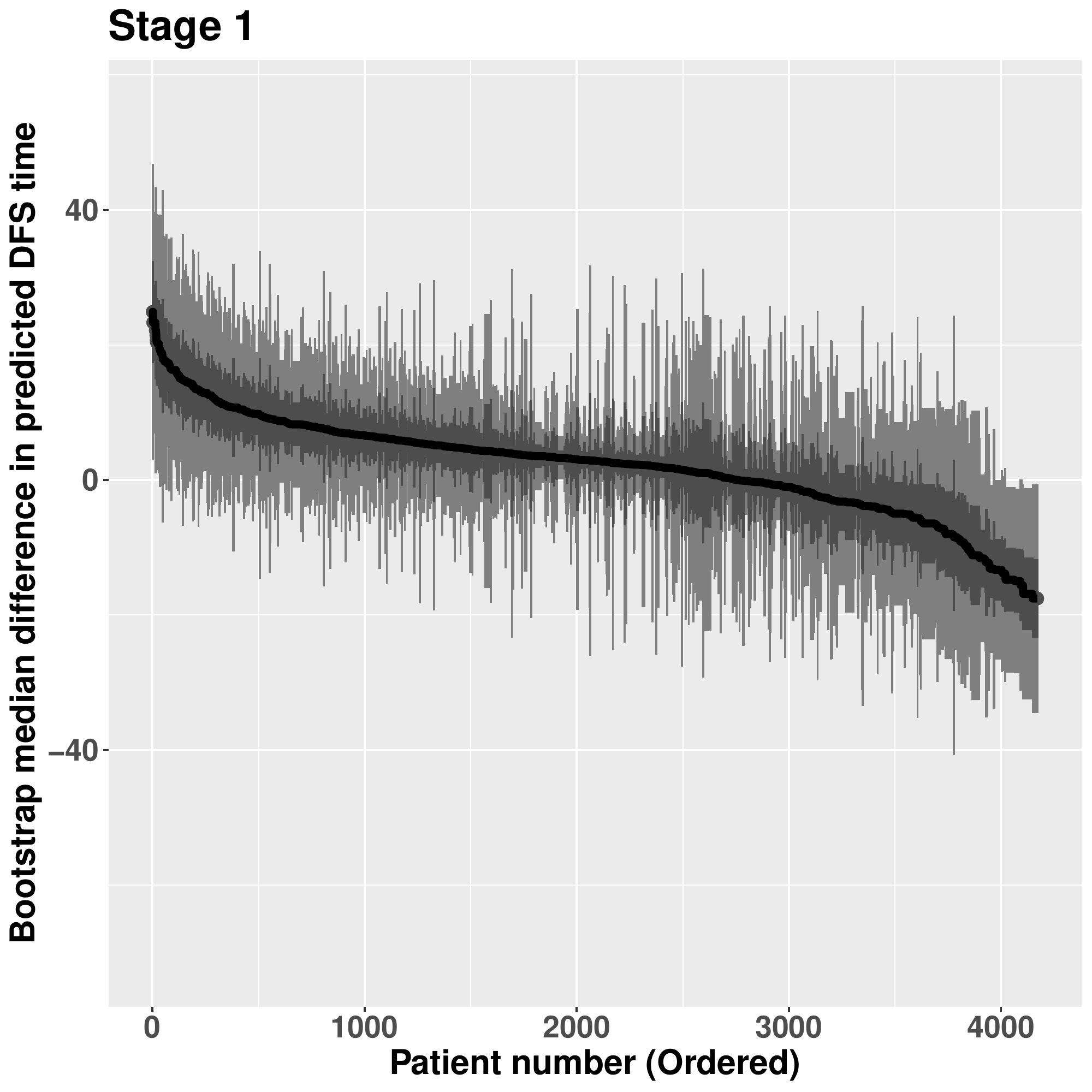}
        \caption[]%
        {{\small Q-learning (NHTL - Standard)}}
        \label{fig:qlearn1dfs}
    \end{subfigure}
    \vskip\baselineskip
    \begin{subfigure}[b]{0.475\textwidth}
        \centering
        \includegraphics[width=\textwidth]{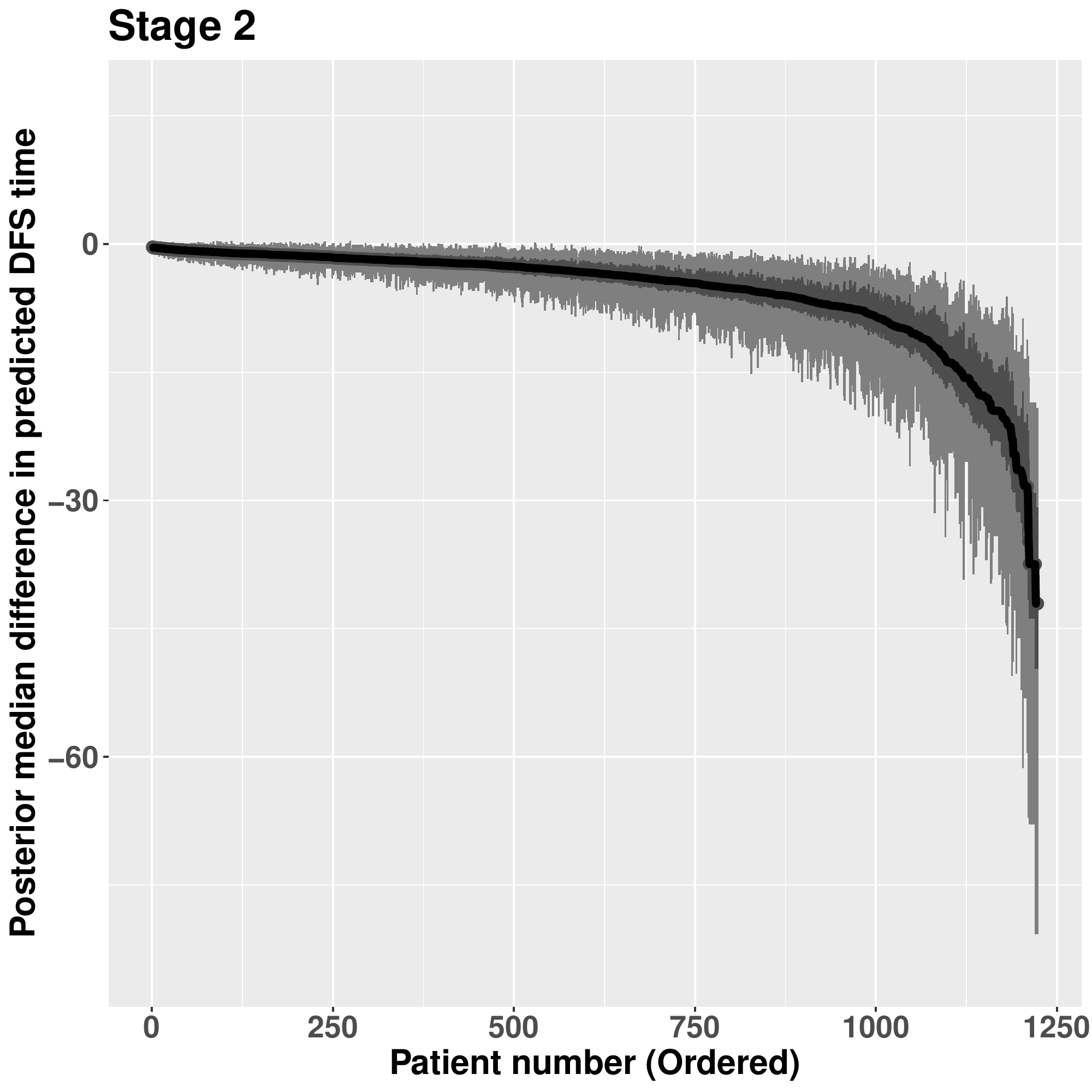}
        \caption[]%
        {{\small AFT-BML (NHTL - Standard)}}
        \label{fig:bart2dfs}
    \end{subfigure}
    \hfill
    \begin{subfigure}[b]{0.475\textwidth}
        \centering
        \includegraphics[width=\textwidth]{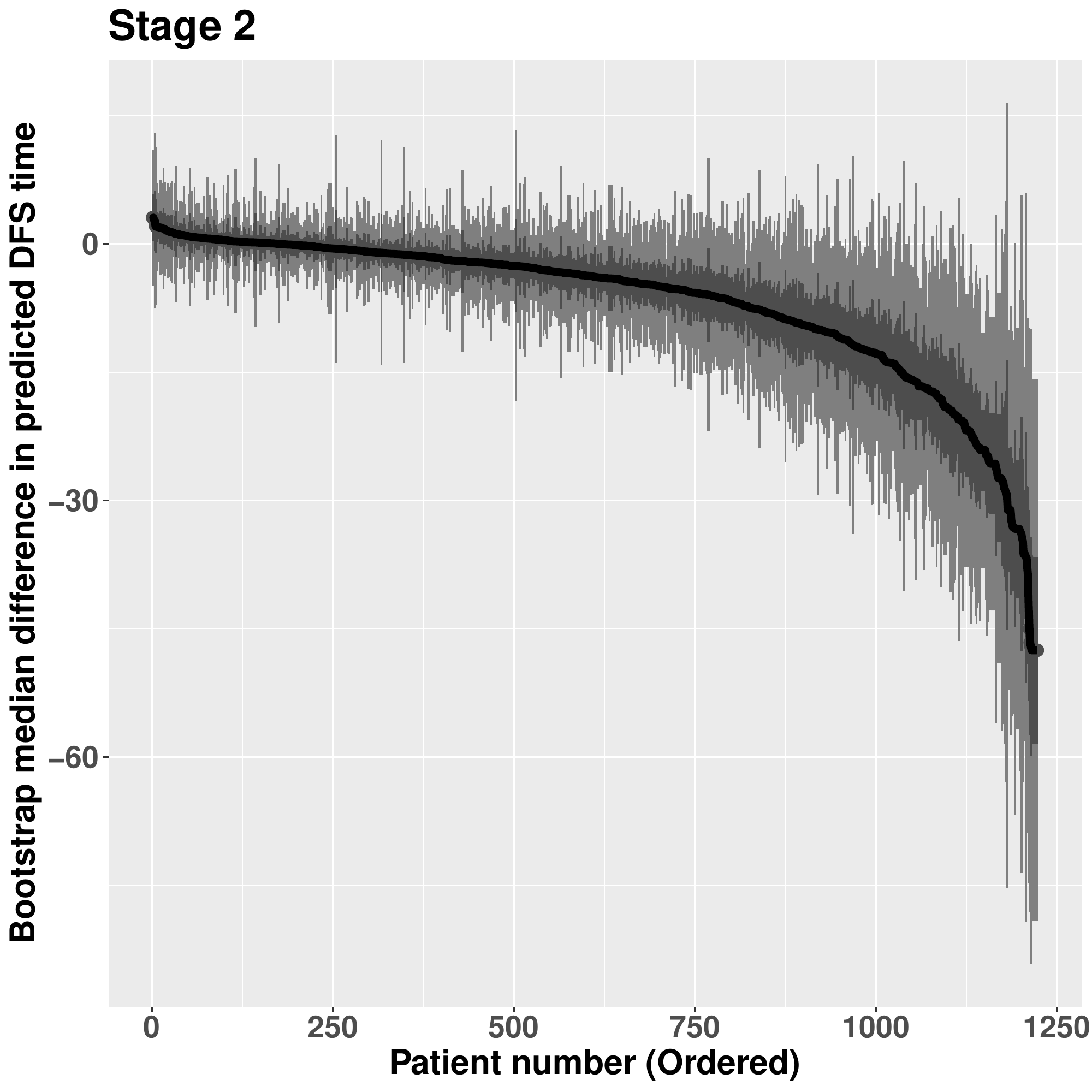}
        \caption[]%
        {{\small Q-learning (NHTL - Standard)}}
        \label{fig:qlearn2dfs}
    \end{subfigure}
    \caption[ ]
    {\small Predicted  difference in median DFS time among patients who received AHCT with NHTL versus standard treatment}
    \label{fig:dfs}
\end{figure*}
In figure \ref{fig:dfs}, we present the predicted treatment differences on a different time scale, specifically the estimated differences of median DFS time with 95\% CI and 50\% CI at each stage. Since the event time is assumed to follow a log normal distribution, the exponential of the log normal mean, which transforms the predictions back to the original time scale, is the median rather than the mean. The results on this scale are generally consistent with the log time scale results, though the scale change produces some unusual results.  There are some patients in the middle of figure \ref{fig:bart1dfs} who have wider CIs than their neighbors. These are patients whose survival predictions are higher (under both treatments), and whose corresponding prediction variances are also higher, leading to a wider CI. Similar observations can be seen on both ends of the same figure, where the patients who are predicted to live longer on either treatment have wider CIs as well.

\begin{figure}[ht]
    \centering
    \begin{subfigure}[b]{0.475\textwidth}
        \centering
        \includegraphics[width=\textwidth]{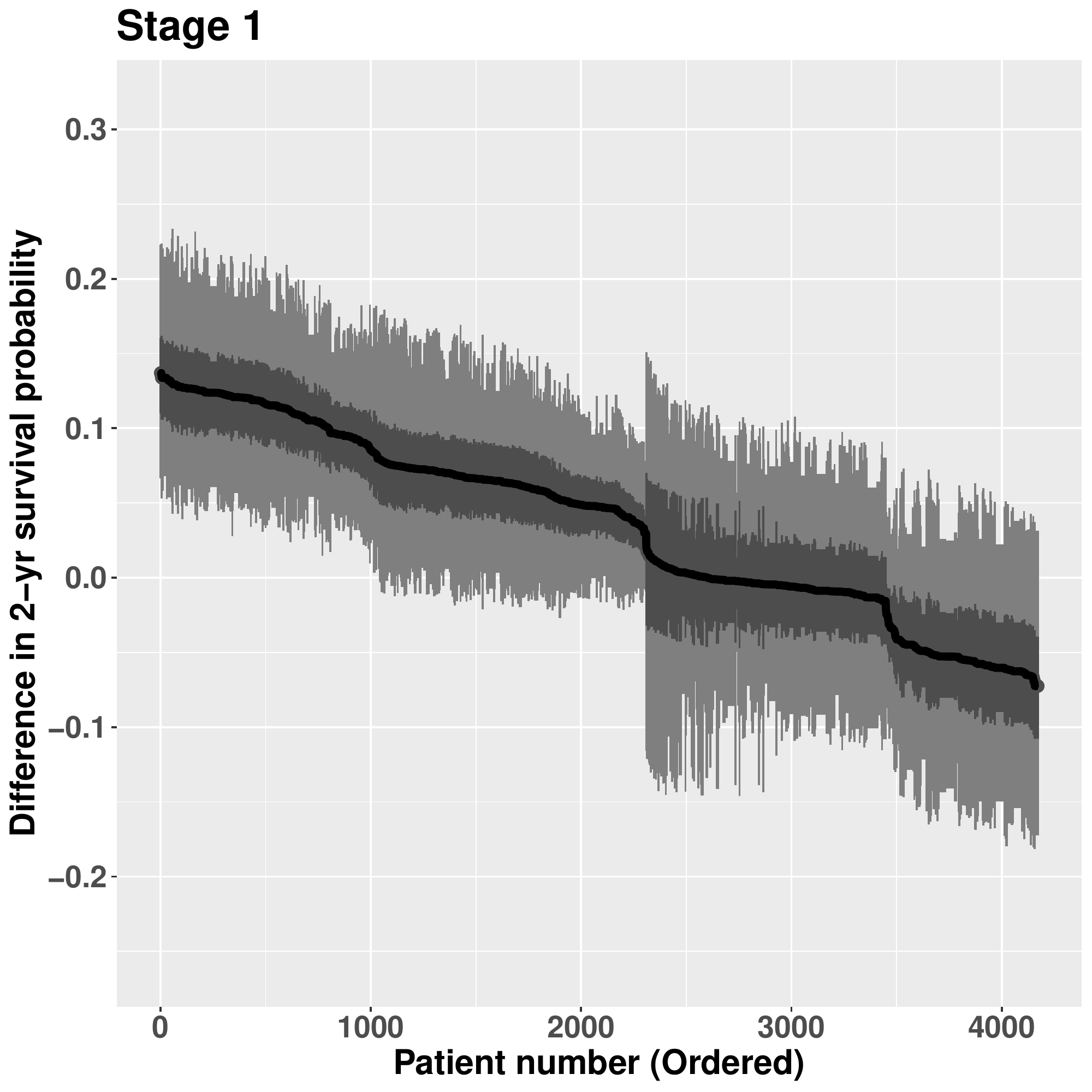}
        \caption[Network2]%
        {{\small AFT-BML (NHTL - Standard)}}
        \label{fig:bart1_2yr}
    \end{subfigure}
    \hfill
    \begin{subfigure}[b]{0.475\textwidth}
        \centering
        \includegraphics[width=\textwidth]{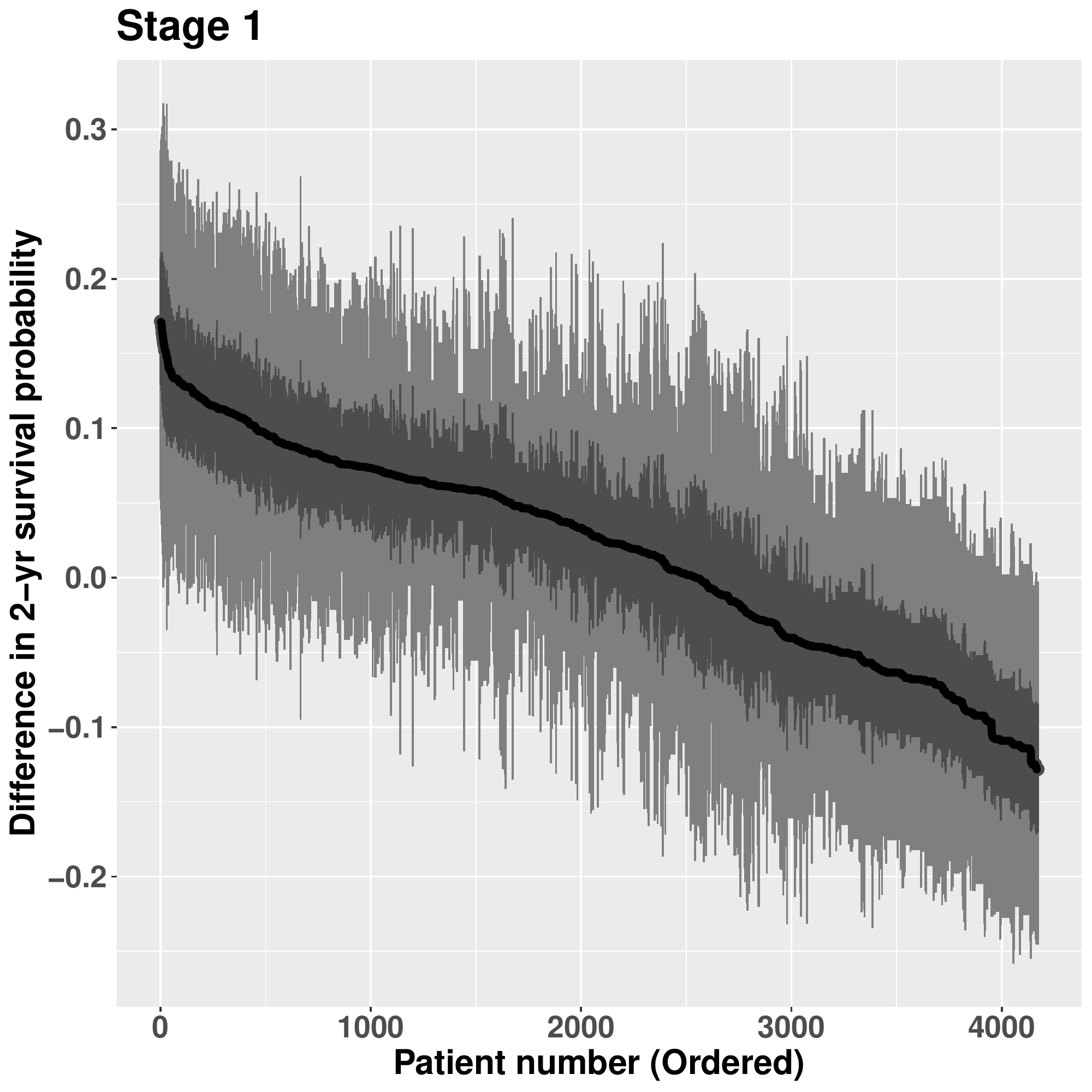}
        \caption[]%
        {{\small Q-learning (NHTL - Standard)}}
        \label{fig:qlearn1_2yr}
    \end{subfigure}
    \vskip\baselineskip
    \begin{subfigure}[b]{0.475\textwidth}
        \centering
        \includegraphics[width=\textwidth]{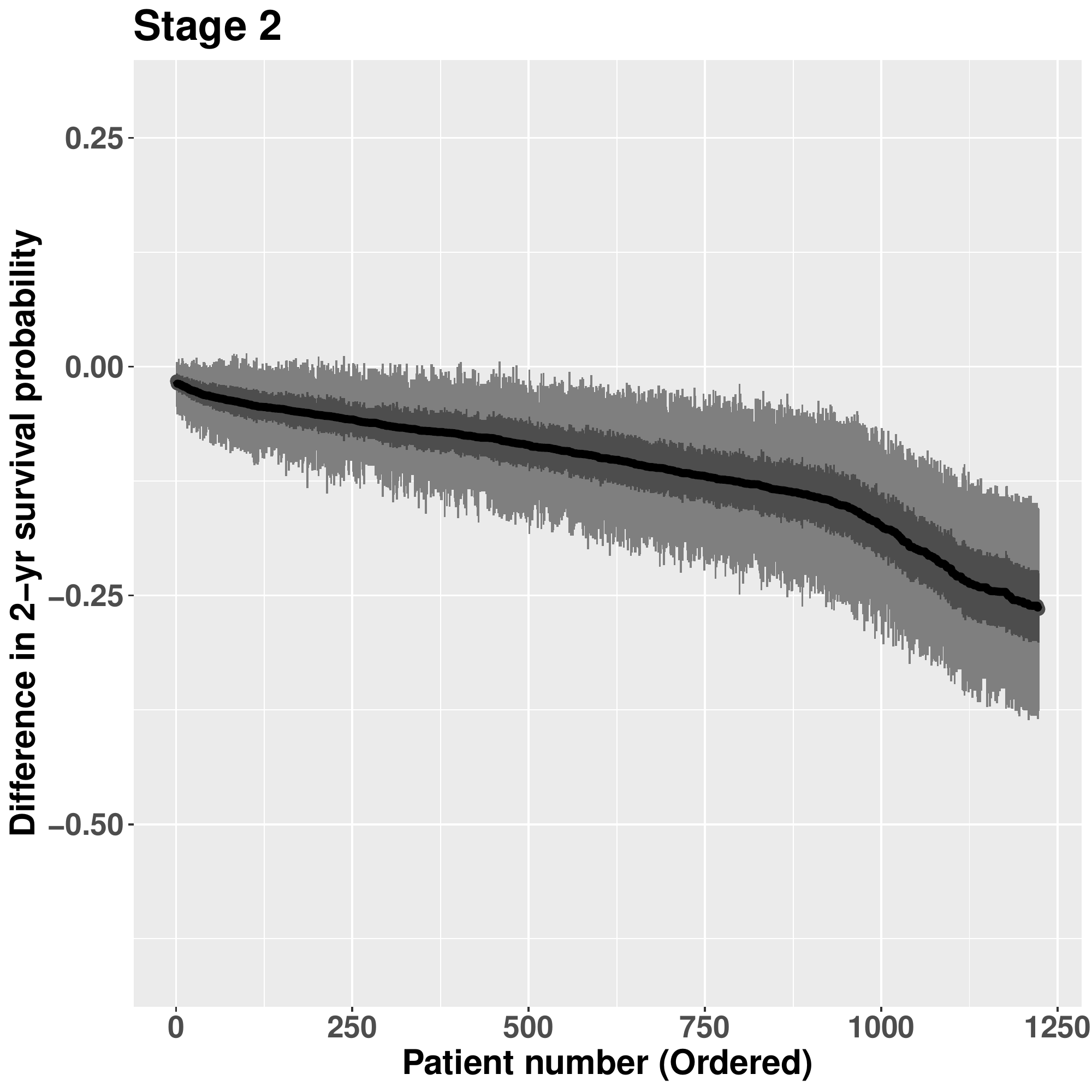}
        \caption[]%
        {{\small AFT-BML (NHTL - Standard)}}
        \label{fig:bart2_2yr}
    \end{subfigure}
    \hfill
    \begin{subfigure}[b]{0.475\textwidth}
        \centering
        \includegraphics[width=\textwidth]{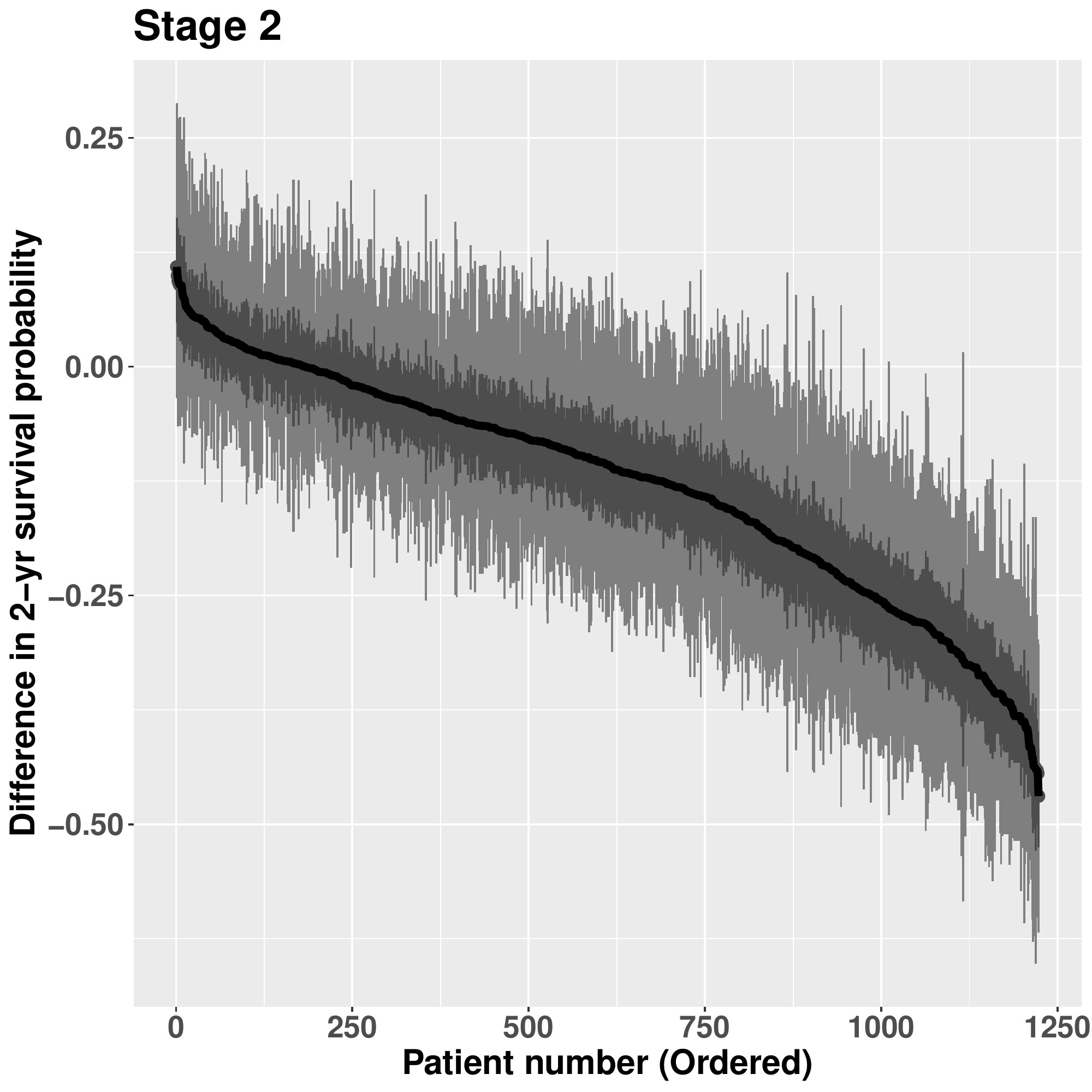}
        \caption[]%
        {{\small Q-learning (NHTL - Standard)}}
        \label{fig:qlearn2_2yr}
    \end{subfigure}
    \caption{2-year DFS probability difference among patients who received AHCT with NHTL versus standard treatment.}
    \label{fig:2yr}
\end{figure}
DTRs could also be defined as the optimal treatment rules to maximize each patient's 2-year DFS probability. Figure \ref{fig:2yr} shows the 2-year DFS probability difference between NHTL and standard treatment for all patients sorted in descending order. For Stage 2, all patients are expected to have a higher 2-year survival probability if assigned the standard treatment based on AFT-BML method. The Q-learning method agrees with the inferences from AFT-BML except for a very small proportion of patients, although the magnitude of the differences are much bigger for Q-learning. For Stage 1, it is easy to notice that AFT-BML splits the patients into four subgroups, similar to what we saw in figure \ref{fig:bart1dfslg}. With Q-learning, however, it is difficult to recognize any clear cut points on the curve. As in Stage 2, the magnitude of the differences are smaller for AFT-BML. The proportions of patients who should have NHTL as the optimal treatment for Stage 1 are consistent between AFT-BML and Q-learning.

To further compare the survival probability predictive performance of AFT-BML vs.\ Q-learning, we calculate the time dependent area under the ROC curve  \cite{heagerty2000time} with the R package \texttt{timeROC}  \cite{timeroc} using the predicted survival time estimated by both AFT-BML and Q-learning as predictors. For the Stage 2 model, the observed time and event indicator can be used directly in calculating the time dependent AUC. For the Stage 1 prediction model (which assumes patients receive optimal treatment in Stage 2), we need to account for not all patients receiving their optimal treatment in Stage 2.  To handle this, depending on whether the estimated optimal treatment at Stage 2 was observed or not, the original observation was kept as is, or was censored at the time of entering Stage 2. Since the estimated optimal Stage 2 treatment could be different from AFT-BML to Q-learning, we examined three sets of censored Stage 1 data, including optimal Stage 2 treatment identified by AFT-BML, or Q-learning, or consistently optimal under both Stage 2 models. The time points of interest for Stage 1 are one year, two years, and three years. For Stage 2, only the median and third quartile of the observed time are evaluated.
%
The results from both stages are shown in Table \ref{tab:aucstg1}.  For Stage 2, AFT-BML improves the AUC by $0.55\%$ at the median, and $1.87\%$ at the third quartile, indicating that AFT-BML has a better predictive performance at Stage 2.
\begin{table}[ht]
    \centering
    \caption{Time dependent AUC for Stage 1 and Stage 2 with either AFT-BML model or Q-learning model.  For Stage 1, observations were censored at entry to Stage 2 for calculation of the time dependent AUC if they did not receive optimal treatment in Stage 2 (with optimal treatment determined using AFT-BML, using Q-learning, or when both agreed).}
    \begin{adjustbox}{width=\textwidth}

    \begin{tabular}{c|c|c|cc}
        &Time in & Suboptimal treatment & \multicolumn{2}{c}{Time dependent AUC} \\
        Stage & months & Censoring Rule & AFT-BML & Q-learning \\
        \hline
        1&12 &AFT-BML based & 71.34\% & 69.61\% \\
        &&Q-learning based&70.89\% & 69.14\% \\
        &&Both agreed&70.81\% & 69.23\% \\
&24&AFT-BML based & 72.68\% & 70.52\% \\
&&Q-learning based & 72.07\% & 69.86\% \\
&&Both agreed & 71.99\% & 69.99\% \\
    &    36 &AFT-BML based & 72.50\% & 70.35\% \\
    && Q-learning based & 71.82\% & 69.60\% \\
    && Both agreed &  71.79\% & 69.86\% \\ \hline
        2&3.2 (Median) & NA & 70.33\% & 69.78\% \\
        &15 (Third quartile) & NA & 76.09\% & 74.22\% \\
    \end{tabular}

    \end{adjustbox}
    \label{tab:aucstg1}
\end{table}
For Stage 1, the time dependent AUC at 1 year from AFT-BML is approximately $1.7\%$ higher than Q-learning in all three settings. This improvement increases to around $2.2\%$ as time goes to 2 years and 3 years. It indicates that AFT-BML once again outshines Q-learning at Stage 1 in predictive performance.

\begin{figure}[h]
    \centering
    \includegraphics[clip, trim={0.7cm 2.7cm 0.6cm 0}, width=\textwidth]{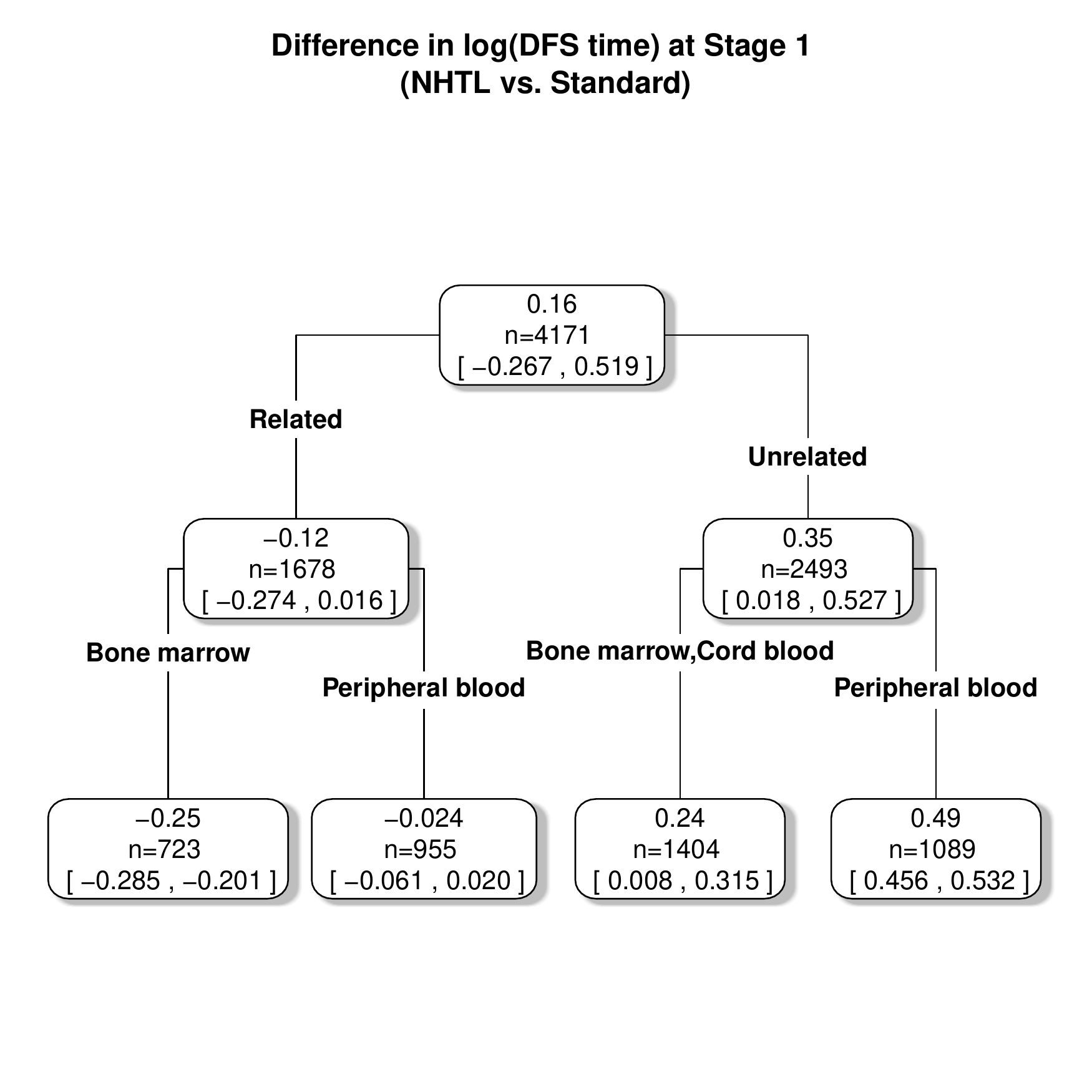}
    \caption{Single tree fit to the posterior mean log(DFS time) of treatment differences as estimated by AFT-BML in Stage 1.  The first split is on the donor relationship (related vs. unrelated), while the second split is on graft source (Bone marrow vs. Peripheral Blood vs. Cord blood).  }
    \label{fig:ftflg}
\end{figure}

To visualize the AFT-BML based DTRs, we applied the `fit-the-fit' method and plotted a single tree as in Logan et al.  \cite{logan2019decision}.  Since there is little value in differentiating the treatment for Stage 2, we only focus on the Stage 1 model here. Here the outcome used for the single tree fit is the posterior mean treatment difference of the log survival time, although other outcomes such as median DFS or DFS probabilities at fixed timepoints could also be used as outcomes. The $R^2$ goodness of fit measure for using a single tree in figure \ref{fig:ftflg} to model the Stage 1 posterior mean differences in log mean DFS time predictions is above 90\%, indicating that this (highly interpretable) single tree is a reasonable representation of the original AFT-BML model at Stage 1. Values in the nodes are the posterior mean differences in mean log DFS time in months between NHTL and standard treatment. The corresponding 95\% CIs are also shown in the same node. The first split (on donor type) indicates that almost all patients receiving unrelated donor transplants would benefit from receiving NHTL as GVHD prophylaxis for their AHCT, while relatively few patients receiving related donor transplants should receive NHTL. As the tree grows, the patients can be divided into four subgroups.  After the first split, the two bottom nodes on the left are well apart from each other, as are the two nodes on the right side. These observations agree with \ref{fig:bart1dfslg}, that identified four subgroups with a distinct posterior mean difference in log DFS times.

\section{Conclusion}
\label{sec:conc}
In medical practice, it is of important clinical value to select the optimal treatment based on the patient's characteristics.  In many settings, a sequence of optimal treatments is desired, such as for diseases like cancer that can progress. The AFT-BML approach for identifying the DTRs can assist physicians to make sound, data-supported decisions at each stage. Classical parametric approaches, such as Q-learning, are constrained by the necessity of correctly specifying functional forms, including all the interaction terms among covariates and treatment. The Bayesian machine learning approach, in contrast, avoids this restriction and allows the model to adaptively determine the relationships between the outcome and the covariates, facilitating optimal treatment identification. We have extended the Bayesian machine learning approach to censored survival data in an AFT model framework, and also provide parallelizable code for implementation of the computationally intensive algorithm.  This wrapper function utilizes standard available BART software without needing to modify the complex underlying BART code.  With the simulation studies, we have shown that the AFT-BML approach can achieve almost the same performance as the oracle model for censored outcomes. The results from AFT-BML not only include the optimal treatment and optimal outcomes that classical parametric approaches can provide, but also directly offers the uncertainty measurement for these targets of inference. This extra uncertainty information could be useful in practice since physicians could better assess the confidence they should have in the recommended optimal treatments.

Our comparison to Q-learning in the simulations and the example used a fixed model specification without variable selection. With a larger number of covariates, a penalized AFT model could be used for variable selection in the Q-learning approach. Furthermore, we utilized bootstrapping to estimate the uncertainty of the Q-learning predictions. Alternative approaches, such as penalized Q-learning  \cite{SongEtAl2014}, could be used to directly provide the uncertainty measurements. However, we are not aware that this approach has been extended to censored data models, such as the AFT model used here.

We compared model performance between AFT-BML and Q-learning in the example by censoring the patients who did not receive optimal subsequent treatment. This was feasible in this data set because most of the patients received optimal treatment in Stage 2. However, a general strategy of assessing the model performance in the dynamic treatment regimes setting would be worth further investigation.

There are some limitations in our approach. We have demonstrated the BML approach for censored data using a parametric log normal AFT model, which has substantial parametric model assumptions even though the functional form of the covariate effects is flexible; other types of Bayesian survival models  \cite{sparapani2016nonparametric,henderson2020individualized,linero2021bayesian}, could alternatively be used which require fewer assumptions.  In such cases, the methodology described here and our wrapper function can serve as a template for implementing alternative models in a BML DTR framework.

Another limitation is that our AFT-BML approach and wrapper function are currently implemented for the two-stage AFT-BART survival model.  Modifying our current approach to handle more than two stages would introduce some computational challenges.  The computational time would increase since we will have more layers of chain burn-in if more than 2 stages are present. This limitation could be reduced if the multi stage model fittings were embedded together, though this would lose the flexibility of applying BART functions off-the-shelf.  Essentially one would need to output the tree structure and terminal node means at the end of one update, and then pass these to the next imputed dataset being analyzed as an initial tree structure and terminal node mean that is being updated.

\section{Acknowledgements*}
This work builds on the M.Sc.\ thesis of S M Ferdous Hossain \cite{FerdousThesis}. The authors are grateful to Dr Elizabeth Krakow for permission to use the analytic dataset created by her for \cite{krakow2017tools}.
The dataset was provided by the Center for International Blood and Marrow Transplant Research (CIBMTR); the CIBMTR data registry is partially supported by the US National Cancer Institute, Grant/Award Number: U24CA076518 and HHSH234200637015C (HRSA/DHHS).

BRL is partially supported by the US National Cancer Institute, Grant/Award Number: U24CA076518 and HHSH234200637015C (HRSA/DHHS).  SMFH was supported by EEMM's Discovery Grant from the Canadian Natural Sciences and Engineering Research Council (NSERC). EEMM is a Canada Research Chair (Tier 1) in Statistical Methods for Precision Medicine and acknowledges the support of a chercheur de m\'erite career award from the Fonds de Recherche du Qu\'ebec, Sant\'e.

 \bibliographystyle{chicago}
\bibliography{Bibliography-MM-MC}

\newpage
\appendix
\section{Analysis results from Q-learning}

\begin{table}[ht]
    \centering
    \caption{Predictors of DFS among patients who received AHCT, according to prophylactic GVHD treatment (A1.NHTL) received. Average estimates from 1000 bootstrap samples and bootstrap confidence intervals (95\% CI) are given.}
    \begin{adjustbox}{width=\textwidth}
    \begin{tabular}{lrrc}
    \hline
        Parameter & Point estimate & Bootstrap mean & 95\% CI \\
         \hline
        (Intercept) &	$2.744$ &	$2.872$ &	$[2.590,	3.256]$ \\
        A1.NHTL &	$-0.257$ &	$-0.404$ &	$[-0.865,	0.063]$ \\
        Age &	$-0.068$ &	$-0.092$ &	$[-0.164,	-0.023]$ \\
        Karnofsky score &	$-0.521$ &	$-0.531$ &	$[-0.700,	-0.348]$ \\
        Disease risk - Early &	$0.741$ &	$0.876$ &	$[0.738,	1.035]$ \\
        Disease risk - Intermediate &	$0.684$ &	$0.685$ &	$[0.512,	0.872]$ \\
        Unrelated donor &	$-0.541$ &	$-0.490$ &	$[-0.750,	-0.263]$ \\
        Female-to-Male &	$-0.049$ &	$0.014$ &	$[-0.168,	0.168]$ \\
        Graftype - Cord blood &	$0.592$ &	$0.415$ &	$[-0.156,	1.014]$ \\
        Graftype - Peripheral blood &	$-0.131$ &	$-0.136$ &	$[-0.357,	0.078]$ \\
        rec\_match &	$-0.087$ &	$-0.139$ &	$[-0.275,	-0.025]$ \\
        TBI &	$-0.018$ &	$-0.105$ &	$[-0.233,	0.025]$ \\
        CMV pair - Negative-Negative &	$0.052$ &	$0.051$ &	$[-0.115,	0.218]$ \\
        Conditioning - RIC\_NMA &	$-0.187$ &	$-0.108$ &	$[-0.371,	0.151]$ \\
        pgvhcor Yes &	$-0.196$ &	$-0.150$ &	$[-0.311,	0.013]$ \\
        A1.NHTL:Age &	$0.019$ &	$-0.008$ &	$[-0.160,	0.143]$ \\
        A1.NHTL:Karnofsky score &	$0.107$ &	$0.173$ &	$[-0.181,	0.549]$ \\
        A1.NHTL:Disease risk - Early &	$0.084$ &	$-0.063$ &	$[-0.356,	0.280]$ \\
        A1.NHTL:Disease risk - Intermediate &	$-0.112$ &	$-0.126$ &	$[-0.422,	0.167]$ \\
        A1.NHTL:Unrelated donor &	$0.398$ &	$0.548$ &	$[0.204,	0.903]$ \\
        A1.NHTL:Female-to-Male &	$-0.016$ &	$0.002$ &	$[-0.328,	0.301]$ \\
        A1.NHTL:Graftype - Cord blood &	$-0.299$ &	$-0.251$ &	$[-0.952,	0.407]$ \\
        A1.NHTL:Graftype - Peripheral blood &	$0.115$ &	$0.198$ &	$[-0.111,	0.531]$ \\
        A1.NHTL:rec\_match &	$-0.137$ &	$0.003$ &	$[-0.177,	0.203]$ \\
        A1.NHTL:TBI &	$-0.094$ &	$0.074$ &	$[-0.211,	0.356]$ \\
        A1.NHTL:CMV pair - Negative-Negative &	$0.313$ &	$0.221$ &	$[-0.063,	0.552]$ \\
        A1.NHTL:Conditioning - RIC\_NMA &	$0.170$ &	$-0.005$ &	$[-0.350,	0.341]$ \\
        A1.NHTL:pgvhcor Yes &	$0.312$ &	$0.162$ &	$[-0.200,	0.507]$ \\
        Log(scale) &	$0.466$ &	$0.510$ &	$[0.442,	0.660]$ \\
         \hline
    \end{tabular}
    \end{adjustbox}
    \label{tab:qstg1}
\end{table}

\begin{table}[ht]
    \centering
    \caption{Predictors of DFS among patients who received AHCT and then proceeded to acute GVHD salvage (A2.NHTL) treatment. Average estimates from 1000 bootstrap samples and bootstrap confidence intervals (95\% CI) are given.}
    \begin{adjustbox}{width=\textwidth}
    \begin{tabular}{lrrc}
    \hline
        Parameter & Point estimate & Bootstrap mean & 95\% CI \\
        \hline
        (Intercept) &	$1.898$ &	$1.727$ &	$[1.171,	2.300]$ \\
        A2.NHTL &	$-1.036$ &	$-0.745$ &	$[-1.761,	0.290]$ \\
        Age &	$-0.244$ &	$-0.034$ &	$[-0.205,	0.140]$ \\
        Karnofsky score &	$-1.015$ &	$-0.858$ &	$[-1.329,	-0.399]$ \\
        Disease risk - Early &	$1.175$ &	$1.051$ &	$[0.711,	1.401]$ \\
        Disease risk - Intermediate &	$0.661$ &	$0.842$ &	$[0.453,	1.241]$ \\
        Unrelated donor &	$-0.577$ &	$-0.799$ &	$[-1.159,	-0.419]$ \\
        Female-to-Male &	$0.060$ &	$-0.074$ &	$[-0.407,	0.292]$ \\
        Graftype - Cord blood &	$0.553$ &	$0.738$ &	$[0.016,	1.505]$ \\
        Graftype - Peripheral blood &	$-0.137$ &	$-0.225$ &	$[-0.563,	0.116]$ \\
        rec\_match &	$-0.290$ &	$-0.246$ &	$[-0.488,	0.000]$ \\
        TBI &	$0.213$ &	$0.354$ &	$[0.044,	0.674]$ \\
        CMV pair - Negative-Negative &	$0.142$ &	$0.377$ &	$[0.046,	0.713]$ \\
        Conditioning - RIC\_NMA &	$-0.112$ &	$-0.025$ &	$[-0.393,	0.336]$ \\
        pgvhcorYes &	$-0.466$ &	$-0.325$ &	$[-0.699,	0.055]$ \\
        Four or more ISP &	$-0.912$ &	$-0.606$ &	$[-0.945,	-0.265]$ \\
        Time to acute GVHD &	$0.406$ &	$0.487$ &	$[0.157,	0.837]$ \\
        A1.NHTL &	$0.670$ &	$0.557$ &	$[0.177,	0.934]$ \\
        A2.NHTL:Age &	$0.108$ &	$-0.217$ &	$[-0.495,	0.063]$ \\
        A2.NHTL:Karnofsky score &	$1.107$ &	$0.563$ &	$[-0.085,	1.232]$ \\
        A2.NHTL:Disease risk - Early &	$-0.683$ &	$-0.665$ &	$[-1.216,	-0.081]$ \\
        A2.NHTL:Disease risk - Intermediate &	$-0.275$ &	$-0.353$ &	$[-1.025,	0.283]$ \\
        A2.NHTL:Unrelated donor &	$0.694$ &	$1.085$ &	$[0.486,	1.673]$ \\
        A2.NHTL:Female-to-Male &	$0.047$ &	$0.235$ &	$[-0.339,	0.841]$ \\
        A2.NHTL:Graftype - Cord blood &	$-0.405$ &	$-0.988$ &	$[-2.486,	0.423]$ \\
        A2.NHTL:Graftype - Peripheral blood &	$0.023$ &	$0.123$ &	$[-0.471,	0.693]$ \\
        A2.NHTL:rec\_match &	$-0.181$ &	$-0.089$ &	$[-0.484,	0.292]$ \\
        A2.NHTL:TBI &	$0.205$ &	$-0.022$ &	$[-0.533,	0.506]$ \\
        A2.NHTL:CMV pair - Negative-Negative &	$0.171$ &	$-0.269$ &	$[-0.824,	0.291]$ \\
        A2.NHTL:Conditioning - RIC\_NMA &	$0.523$ &	$0.213$ &	$[-0.485,	0.907]$ \\
        A2.NHTL:pgvhcorYes &	$-0.167$ &	$-0.131$ &	$[-0.733,	0.422]$ \\
        A2.NHTL:Four or more ISP &	$0.559$ &	$0.358$ &	$[-0.199,	0.911]$ \\
        A2.NHTL:Time to acute GVHD &	$-0.169$ &	$-0.324$ &	$[-0.900,	0.276]$ \\
        A2.NHTL:A1.NHTL &	$-0.649$ &	$-0.596$ &	$[-1.163,	-0.047]$ \\
        Log(scale) &	$0.661$ &	$0.693$ &	$[0.635,	0.749]$ \\
         \hline
    \end{tabular}
    \end{adjustbox}
    \label{tab:qstg2}
\end{table}

\newpage
\section{Software}
\label{appen:func}
The R wrapper function that we created for this chapter can be found at \url{https://github.com/xiaoli-mcw/dtrBART}. Simply download and save the \texttt{wrapper.R} file. Note that this function utilizes some functions from the \textbf{BART3} R package, which is not available on CRAN yet. The \textbf{BART3} package can be found at \url{https://github.com/rsparapa/bnptools}. The function \texttt{dtr1} can be called conveniently after sourcing \texttt{wrapper.R} using
\begin{verbatim}
source("myfolder/wrapper.R")
\end{verbatim}
Remember to replace \texttt{myfolder} with the actual directory where \texttt{wrapper.R} is saved.

The data should have all the covariates including treatments observed in both stages, an indicator for entering Stage 2 defined as $\eta_i$ in Section \ref{sec:aftbml}, an overall event indicator $\delta_i$ regardless of entering Stage 2 or not, a survival time for Stage 1, and a survival time for Stage 2. For those who did not enter Stage 2, since their information are not used in fitting Stage 2 model, their survival time for Stage 2 can be coded in any reasonable way.

\begin{verbatim}
dtr1(x1=c("x1"), a1="a1", time1="y1",
    x2=c("x1","x2"), a2="a2", stg2ind="eta",
    time2="y2", delta="delta", data,
    newdata=NULL, opt=TRUE, mc.cores=8)
\end{verbatim}
In this function, \texttt{x1} is a vector of covariate names that are used in fitting the Stage 1 model. \texttt{a1} is the variable name of the action in Stage 1. \texttt{time1} is the variable name of survival time at Stage 1. \texttt{x2}, \texttt{a2}, and \texttt{time2} are similar for Stage 2. \texttt{stg2ind} is the variable name indicating whether an individual entered Stage 2 indicator. \texttt{delta} is the variable name of the overall event indicator. \texttt{data} is the data set. If a \texttt{newdata} is provided, predictions of optimal action and optimal outcome for the new data will be returned. \texttt{opt} is \texttt{TRUE} or \texttt{FALSE}, indicating whether only the optimal action and survival time at each stage will be returned (\texttt{TRUE}) or whether additional survival times under each action option at each stage will be returned (\texttt{FALSE}). \texttt{mc.cores} specifies the number of threads to be used in the calculation. With more threads, the program will execute more quickly.

There is a demo data set \texttt{dtrdata.csv} in the same repository on GitHub which is used to explain the results  returned by our wrapper function. Here, the \texttt{dtrdata} with 1000 observations is split into training ($80\%$) and testing ($20\%$) data.
\begin{verbatim}
> ind <- sample(1:n, 800)
> train <- dtrdata[ind,]
> test <- dtrdata[-ind,]
> res.dtr <- dtr1(x1="x1", a1="a1", time1="t1",
                x2="x2", a2="a2", time2="t2",
                stg2ind="eta", delta="delta",
                data=train, newdata=test,
                opt=FALSE)
> str(res.dtr)
List of 14
 $ a2.opt         : int [1:476, 1:1000] 1 1 1 0
                        1 0 0 1 0 0 ...
 $ yhat2optmean   : num [1:476, 1:1000] 4.82 4.88
                        5.06 4.5 4.7 ...
 $ newa2.opt      : int [1:121, 1:1000] 0 0 0 1
                        0 0 1 1 0 1 ...
 $ newyhat2optmean: num [1:121, 1:1000] 4.43 4.47
                        4.33 4.71 4.65 ...
 $ sigma2         : num [1:1000] 0.323 0.331
                        0.321 0.326 0.333 ...
 $ a1.opt         : num [1:800, 1:1000] 0 0 1 0
                        1 0 1 1 1 1 ...
 $ yhat1optmean   : num [1:800, 1:1000] 6.54 6.4
                        7.26 6.38 7.32 ...
 $ newa1.opt      : num [1:200, 1:1000] 1 1 1 1
                        1 1 0 0 1 1 ...
 $ newyhat1optmean: num [1:200, 1:1000] 6.99 6.83
                        7.42 6.82 7.19 ...
 $ sigma1         : num [1:1000] 0.335 0.348
                        0.345 0.346 0.342 ...
 $ a2_0           : num [1:597, 1:1000] 4.64 4.72
                        4.79 4.5 4.62 ...
 $ a2_1           : num [1:597, 1:1000] 4.82 4.88
                        5.06 4.37 4.7 ...
 $ a1_0           : num [1:800, 1:1000] 6.54 6.4
                        7 6.38 7.1 ...
 $ a1_1           : num [1:800, 1:1000] 6.51 6.37
                        7.26 6.36 7.32 ...
\end{verbatim}
The result \texttt{res.dtr} is a list of optimal actions (\texttt{a2.opt}, \texttt{a1.opt}) and corresponding outcomes (\texttt{yhat2optmean}, \texttt{yhat1optmean}) estimated for both stages, as well as the variances of estimated outcome (\texttt{sigma2}, \texttt{sigma1}). Since we supplied testing data to the wrapper function, we have the predicted optimal actions (\texttt{newa2.opt}, \texttt{newa1.opt}) and corresponding outcomes (\texttt{newyhat2optmean}, \texttt{newyhat1optmean}) reported. The estimated outcomes under each possible action at both stages (\texttt{a2\_0}, \texttt{a2\_1}, \texttt{a1\_0}, \texttt{a1\_1}) for the training data are also presented because we set \texttt{opt=FALSE}. The rows represent individuals, and the columns represent the MCMC samples.

\end{document}